\let\KernelLabel\label
\AddToHook{package/hyperref/before}{\let\label\KernelLabel}
\documentclass[twocolumn]{aastex62}
\shorttitle{Extinction Curve Variation in DESI}
\shortauthors{Singh et al.}
\usepackage{textcomp}
\usepackage{hyperref}
\usepackage{amsmath}
\usepackage{amsfonts}
\usepackage{booktabs}
\usepackage{lipsum}
\usepackage{lineno}
\usepackage{multirow}
\usepackage{xspace}
\usepackage{xurl}
\usepackage{microtype}
\usepackage{needspace}
\newcommand{\degree}{\ensuremath{^\circ}}

\newcommand{\feh}{\ensuremath{[{\rm Fe/H}]}}
\newcommand{\logg}{\ensuremath{\log g}}
\newcommand{\teff}{\ensuremath{T_{\rm eff}}}

\newcommand{\ebv}{\ensuremath{E(B-V)_{\rm SFD}}}
\newcommand{\rvs}{\ensuremath{R(V)_{\rm S}}}
\newcommand{\rvz}{\ensuremath{R(V)_{\rm Z}}}
\newcommand{\wise}{\textit{WISE}\xspace}
\newcommand{\planck}{\textit{Planck}\xspace}
\newcommand{\gaia}{\textit{Gaia}\xspace}

\makeatletter
\let\revtex@check@deferlist@stuck\check@deferlist@stuck
\def\check@deferlist@stuck#1{%
  \ifx#1\do@startpage
    \begingroup
      \global\@tempswafalse
      \def\@elt##1{\ifdim\wd##1<\textwidth\else\global\@tempswatrue\fi}%
      \@deferlist
    \endgroup
    \if@tempswa\expandafter\revtex@check@deferlist@stuck\expandafter#1\fi
  \else
    \expandafter\revtex@check@deferlist@stuck\expandafter#1%
  \fi}
\makeatother

\begin{document}
\enlargethispage{-12pt}

\title{Variations in the Extinction Curve at High Galactic Latitudes as seen by the Dark Energy Spectroscopic Instrument}

\author[0009-0005-7982-2369]{Sarv Parteek Singh}
\affiliation{Johns Hopkins University, 3400 N. Charles Street, Baltimore, MD 21218, USA}
\correspondingauthor{Sarv Parteek Singh}
\email{ssingh60@jh.edu}

\author[0000-0002-3569-7421]{Edward~F.~Schlafly}
\affiliation{Space Telescope Science Institute, 3700 San Martin Drive, Baltimore, MD 21218, USA}

\author{Rongpu Zhou}
\affiliation{Lawrence Berkeley National Laboratory, 1 Cyclotron Road, Berkeley, CA 94720, USA}

\author{Sergey~E.~Koposov}
\affiliation{Institute for Astronomy, University of Edinburgh, Royal Observatory, Blackford Hill, Edinburgh EH9 3HJ, UK}
\affiliation{Institute of Astronomy, University of Cambridge, Madingley Road, Cambridge CB3 0HA, UK}

\author{Joan~R.~Najita}
\affiliation{NSF’s NOIRLab, 950 N. Cherry Avenue, Tucson, AZ 85719, USA}

\author{J.~Aguilar}
\affiliation{Lawrence Berkeley National Laboratory, 1 Cyclotron Road, Berkeley, CA 94720, USA}

\author{S.~Ahlen}
\affiliation{Department of Physics, Boston University, 590 Commonwealth Avenue, Boston, MA 02215 USA}

\author{C.~Allende~Prieto}
\affiliation{Departamento de Astrof\'{\i}sica, Universidad de La Laguna (ULL), E-38206, La Laguna, Tenerife, Spain}
\affiliation{Instituto de Astrof\'{\i}sica de Canarias, C/ V\'{\i}a L\'{a}ctea, s/n, E-38205 La Laguna, Tenerife, Spain}

\author{D.~Bianchi}
\affiliation{Dipartimento di Fisica ``Aldo Pontremoli'', Universit\`a degli Studi di Milano, Via Celoria 16, I-20133 Milano, Italy}
\affiliation{INAF-Osservatorio Astronomico di Brera, Via Brera 28, 20122 Milano, Italy}

\author{D.~Brooks}
\affiliation{Department of Physics \& Astronomy, University College London, Gower Street, London, WC1E 6BT, UK}

\author{A.~Carnero Rosell}
\affiliation{Departamento de Astrof\'{\i}sica, Universidad de La Laguna (ULL), E-38206, La Laguna, Tenerife, Spain}
\affiliation{Instituto de Astrof\'{\i}sica de Canarias, C/ V\'{\i}a L\'{a}ctea, s/n, E-38205 La Laguna, Tenerife, Spain}

\author{F.~J.~Castander}
\affiliation{Institut d'Estudis Espacials de Catalunya (IEEC), c/ Esteve Terradas 1, Edifici RDIT, Campus PMT-UPC, 08860 Castelldefels, Spain}
\affiliation{Institute of Space Sciences, ICE-CSIC, Campus UAB, Carrer de Can Magrans s/n, 08913 Bellaterra, Barcelona, Spain}

\author{E.~Chaussidon}
\affiliation{Lawrence Berkeley National Laboratory, 1 Cyclotron Road, Berkeley, CA 94720, USA}

\author{T.~Claybaugh}
\affiliation{Lawrence Berkeley National Laboratory, 1 Cyclotron Road, Berkeley, CA 94720, USA}

\author{A.~P.~Cooper}
\affiliation{Institute of Astronomy and Department of Physics, National Tsing Hua University, 101 Kuang-Fu Rd. Sec. 2, Hsinchu 30013, Taiwan}

\author{A.~Cuceu}
\affiliation{Lawrence Berkeley National Laboratory, 1 Cyclotron Road, Berkeley, CA 94720, USA}

\author{K.~S.~Dawson}
\affiliation{Department of Physics and Astronomy, The University of Utah, 115 South 1400 East, Salt Lake City, UT 84112, USA}

\author{A.~de la Macorra}
\affiliation{Instituto de F\'{\i}sica, Universidad Nacional Aut\'{o}noma de M\'{e}xico,  Circuito de la Investigaci\'{o}n Cient\'{\i}fica, Ciudad Universitaria, Cd. de M\'{e}xico  C.~P.~04510,  M\'{e}xico}

\author{Biprateep~Dey}
\affiliation{Department of Astronomy \& Astrophysics, University of Toronto, Toronto, ON M5S 3H4, Canada}
\affiliation{Department of Physics \& Astronomy and Pittsburgh Particle Physics, Astrophysics, and Cosmology Center (PITT PACC), University of Pittsburgh, 3941 O'Hara Street, Pittsburgh, PA 15260, USA}

\author{P.~Doel}
\affiliation{Department of Physics \& Astronomy, University College London, Gower Street, London, WC1E 6BT, UK}

\author{S.~Ferraro}
\affiliation{Lawrence Berkeley National Laboratory, 1 Cyclotron Road, Berkeley, CA 94720, USA}
\affiliation{University of California, Berkeley, 110 Sproul Hall \#5800 Berkeley, CA 94720, USA}

\author{J.~E.~Forero-Romero}
\affiliation{Departamento de F\'isica, Universidad de los Andes, Cra. 1 No. 18A-10, Edificio Ip, CP 111711, Bogot\'a, Colombia}
\affiliation{Observatorio Astron\'omico, Universidad de los Andes, Cra. 1 No. 18A-10, Edificio H, CP 111711 Bogot\'a, Colombia}

\author{E.~Gaztañaga}
\affiliation{Institut d'Estudis Espacials de Catalunya (IEEC), c/ Esteve Terradas 1, Edifici RDIT, Campus PMT-UPC, 08860 Castelldefels, Spain}
\affiliation{Institute of Cosmology and Gravitation, University of Portsmouth, Dennis Sciama Building, Portsmouth, PO1 3FX, UK}
\affiliation{Institute of Space Sciences, ICE-CSIC, Campus UAB, Carrer de Can Magrans s/n, 08913 Bellaterra, Barcelona, Spain}

\author{Satya~{Gontcho A Gontcho}}
\affiliation{University of Virginia, Department of Astronomy, Charlottesville, VA 22904, USA}

\author{G.~Gutierrez}
\affiliation{Fermi National Accelerator Laboratory, PO Box 500, Batavia, IL 60510, USA}

\author{J.~Guy}
\affiliation{Lawrence Berkeley National Laboratory, 1 Cyclotron Road, Berkeley, CA 94720, USA}

\author{C.~Hahn}
\affiliation{Department of Astronomy, University of Texas at Austin, 2515 Speedway, TX 78712, USA}

\author{H.~K.~Herrera-Alcantar}
\affiliation{Institut d'Astrophysique de Paris. 98 bis boulevard Arago. 75014 Paris, France}
\affiliation{Center for Cosmology and AstroParticle Physics, The Ohio State University, 191 West Woodruff Avenue, Columbus, OH 43210, USA}
\affiliation{Department of Astronomy, The Ohio State University, 4055 McPherson Laboratory, 140 W 18th Avenue, Columbus, OH 43210, USA}
\affiliation{IRFU, CEA, Universit\'{e} Paris-Saclay, F-91191 Gif-sur-Yvette, France}
\affiliation{The Ohio State University, Columbus, 43210 OH, USA}

\author{R.~Joyce}
\affiliation{NSF NOIRLab, 950 N. Cherry Ave., Tucson, AZ 85719, USA}

\author{T.~Karim}
\affiliation{Department of Astronomy \& Astrophysics, University of Toronto, Toronto, ON M5S 3H4, Canada}

\author{R.~Kehoe}
\affiliation{Department of Physics, Southern Methodist University, 3215 Daniel Avenue, Dallas, TX 75275, USA}

\author{D.~Kirkby}
\affiliation{Department of Physics and Astronomy, University of California, Irvine, 92697, USA}

\author{A.~Kremin}
\affiliation{Lawrence Berkeley National Laboratory, 1 Cyclotron Road, Berkeley, CA 94720, USA}

\author{A.~Lambert}
\affiliation{Lawrence Berkeley National Laboratory, 1 Cyclotron Road, Berkeley, CA 94720, USA}

\author{C.~Lamman}
\affiliation{The Ohio State University, Columbus, 43210 OH, USA}

\author{M.~Landriau}
\affiliation{Lawrence Berkeley National Laboratory, 1 Cyclotron Road, Berkeley, CA 94720, USA}

\author{L.~Le~Guillou}
\affiliation{Sorbonne Universit\'{e}, CNRS/IN2P3, Laboratoire de Physique Nucl\'{e}aire et de Hautes Energies (LPNHE), FR-75005 Paris, France}

\author{M.~E.~Levi}
\affiliation{Lawrence Berkeley National Laboratory, 1 Cyclotron Road, Berkeley, CA 94720, USA}

\author{M.~Manera}
\affiliation{Departament de F\'{i}sica, Serra H\'{u}nter, Universitat Aut\`{o}noma de Barcelona, 08193 Bellaterra (Barcelona), Spain}
\affiliation{Institut de F\'{i}sica d’Altes Energies (IFAE), The Barcelona Institute of Science and Technology, Edifici Cn, Campus UAB, 08193, Bellaterra (Barcelona), Spain}

\author{P.~Martini}
\affiliation{Center for Cosmology and AstroParticle Physics, The Ohio State University, 191 West Woodruff Avenue, Columbus, OH 43210, USA}
\affiliation{Department of Astronomy, The Ohio State University, 4055 McPherson Laboratory, 140 W 18th Avenue, Columbus, OH 43210, USA}
\affiliation{The Ohio State University, Columbus, 43210 OH, USA}

\author{A.~Meisner}
\affiliation{NSF NOIRLab, 950 N. Cherry Ave., Tucson, AZ 85719, USA}

\author{R.~Miquel}
\affiliation{Instituci\'{o} Catalana de Recerca i Estudis Avan\c{c}ats, Passeig de Llu\'{\i}s Companys, 23, 08010 Barcelona, Spain}
\affiliation{Institut de F\'{i}sica d’Altes Energies (IFAE), The Barcelona Institute of Science and Technology, Edifici Cn, Campus UAB, 08193, Bellaterra (Barcelona), Spain}

\author{J.~Moustakas}
\affiliation{Department of Physics and Astronomy, Siena University, 515 Loudon Road, Loudonville, NY 12211, USA}

\author{A.~D.~Myers}
\affiliation{Department of Physics \& Astronomy, University  of Wyoming, 1000 E. University, Dept.~3905, Laramie, WY 82071, USA}

\author{S.~Nadathur}
\affiliation{Institute of Cosmology and Gravitation, University of Portsmouth, Dennis Sciama Building, Portsmouth, PO1 3FX, UK}

\author{J.~ A.~Newman}
\affiliation{Department of Physics \& Astronomy and Pittsburgh Particle Physics, Astrophysics, and Cosmology Center (PITT PACC), University of Pittsburgh, 3941 O'Hara Street, Pittsburgh, PA 15260, USA}

\author{E.~Paillas}
\affiliation{Instituto de Estudios Astrof\'isicos, Facultad de Ingenier\'ia y Ciencias, Universidad Diego Portales, Av. Ej\'ercito Libertador 441, Santiago, Chile}

\author{W.~J.~Percival}
\affiliation{Department of Physics and Astronomy, University of Waterloo, 200 University Ave W, Waterloo, ON N2L 3G1, Canada}
\affiliation{Perimeter Institute for Theoretical Physics, 31 Caroline St. North, Waterloo, ON N2L 2Y5, Canada}
\affiliation{Waterloo Centre for Astrophysics, University of Waterloo, 200 University Ave W, Waterloo, ON N2L 3G1, Canada}

\author{C.~Poppett}
\affiliation{Lawrence Berkeley National Laboratory, 1 Cyclotron Road, Berkeley, CA 94720, USA}
\affiliation{Space Sciences Laboratory, University of California, Berkeley, 7 Gauss Way, Berkeley, CA  94720, USA}
\affiliation{University of California, Berkeley, 110 Sproul Hall \#5800 Berkeley, CA 94720, USA}

\author{F.~Prada}
\affiliation{Instituto de Astrof\'{i}sica de Andaluc\'{i}a (CSIC), Glorieta de la Astronom\'{i}a, s/n, E-18008 Granada, Spain}

\author{I.~P\'erez-R\`afols}
\affiliation{Departament de F\'isica, EEBE, Universitat Polit\`ecnica de Catalunya, c/Eduard Maristany 10, 08930 Barcelona, Spain}

\author{C.~Ravoux}
\affiliation{Universit\'{e} Clermont-Auvergne, CNRS, LPCA, 63000 Clermont-Ferrand, France}

\author{G.~Rossi}
\affiliation{Department of Physics and Astronomy, Sejong University, 209 Neungdong-ro, Gwangjin-gu, Seoul 05006, Republic of Korea}

\author{R.~Ruggeri}
\affiliation{Queensland University of Technology,  School of Chemistry \& Physics, George St, Brisbane 4001, Australia}

\author{L.~Samushia}
\affiliation{Abastumani Astrophysical Observatory, Tbilisi, GE-0179, Georgia}
\affiliation{Department of Physics, Kansas State University, 116 Cardwell Hall, Manhattan, KS 66506, USA}

\author{E.~Sanchez}
\affiliation{CIEMAT, Avenida Complutense 40, E-28040 Madrid, Spain}

\author{C.~Saulder}
\affiliation{Max Planck Institute for Extraterrestrial Physics, Gie\ss enbachstra\ss e 1, 85748 Garching, Germany}

\author{D.~Schlegel}
\affiliation{Lawrence Berkeley National Laboratory, 1 Cyclotron Road, Berkeley, CA 94720, USA}

\author{M.~Schubnell}
\affiliation{Department of Physics, University of Michigan, 450 Church Street, Ann Arbor, MI 48109, USA}
\affiliation{University of Michigan, 500 S. State Street, Ann Arbor, MI 48109, USA}

\author{H.~Seo}
\affiliation{Department of Physics \& Astronomy, Ohio University, 139 University Terrace, Athens, OH 45701, USA}

\author{J.~Silber}
\affiliation{Lawrence Berkeley National Laboratory, 1 Cyclotron Road, Berkeley, CA 94720, USA}

\author{M.~Siudek}
\affiliation{Institute of Space Sciences, ICE-CSIC, Campus UAB, Carrer de Can Magrans s/n, 08913 Bellaterra, Barcelona, Spain}
\affiliation{Instituto de Astrof\'{\i}sica de Canarias, C/ V\'{\i}a L\'{a}ctea, s/n, E-38205 La Laguna, Tenerife, Spain}

\author{G.~Tarl\'{e}}
\affiliation{University of Michigan, 500 S. State Street, Ann Arbor, MI 48109, USA}

\author{B.~A.~Weaver}
\affiliation{NSF NOIRLab, 950 N. Cherry Ave., Tucson, AZ 85719, USA}

\author{H.~Zou}
\affiliation{National Astronomical Observatories, Chinese Academy of Sciences, A20 Datun Road, Chaoyang District, Beijing, 100101, P.~R.~China}

\begin{abstract}
The dust extinction curve, characterized by the total-to-selective extinction ratio, $R(V)$, varies with grain size and composition. Accurate $R(V)$ values are crucial for dereddening astronomical colors and understanding the interstellar medium. We measure extinction toward stars with Dark Energy Spectroscopic Instrument (DESI) spectroscopy using a machine-learning-based relative of the ``pair method,'' matching reddened and unreddened stars. We use XGBoost to predict stars' intrinsic colors from their spectroscopic parameters (\teff, \logg, \feh) and subtract them from observed colors to measure reddening in combinations of the $g$, $r$, $z$, $W1$, and $W2$ bands. This approach directly measures reddening, unlike the common thermal-emission-based method of \citealt{SFD_1998} (SFD). The reddening is accurate to roughly 0.02 mag in $g-r$ and $r-z$. We develop a forward model to map $g-r$ and $g-W1$ reddenings to $R(V)$ and find significant variation in $R(V)$, with $\sigma_{\rm{NMAD}}(R(V)) = 0.23$, in dusty regions ($E(g-r) > 0.1$) at high Galactic latitudes observed by DESI. In this footprint, $R(V)$ is bimodal, with modes at 2.9 and 3.4, likely due to large-spatial-scale correlations in the shape of the extinction curve combined with the relatively small number of distinct sightlines with substantial reddening. We find no strong relationship between $R(V)$ and other tracers of the interstellar medium, such as thermal emission, dust temperature, spectral emissivity index $\beta$, and $H\alpha$. However, the error in the SFD-predicted extinction correlates well with $R(V)$, implying that a single optical reddening combined with the far-infrared dust brightness can predict variations in the extinction curve.
\end{abstract}


\section{Introduction} 
\label{sec:intro}

Dust is a major source of systematic uncertainty for many current and future cosmology experiments. For example, by reddening and extinguishing starlight, dust changes the measured distances to supernovae (\citealt{Tripp:1998, Brout&Scolnic:2021})  and imprints spurious structure on the observed distribution of galaxies \citep{Rezaie:2020}.  This effect can limit the accuracy to which the expansion history of the universe can be measured \citep{Betoule:2014}, while scatter in the correction contributes to the noise budget of large-scale clustering measurements \citep{Kitanidis:2020, Zhou_2025}. 

Correcting the effects of dust extinction requires not only accounting for the amount of extinction but also for the shape of the extinction curve \citep{Schlafly:2010, Schlafly:2011}, which varies across the sky and encodes the physical properties of dust \citep{Draine:2003}.

Today, corrections for the effects of dust extinction are most often performed using measurements of the thermal brightness of dust (e.g., from \planck; \citealt{Planck:2014}) or of the intensity of the 21~cm line emission from neutral hydrogen \citep{H14PI:2016, Lenz:2017}.  The conversion of these measurements into dust extinction depends on quantities like the dust temperature, the extinction curve, and the dust-to-gas ratio. Systematic errors in those quantities now limit the accuracy to which dust corrections can be made \citep{Schlafly:2011, Zhou_2025}. Improved correction for dust extinction must be made by other approaches.

Fortunately, direct extinction measurements via the ``pair method'' offer a promising alternative to these indirect measurements. In the pair method, the spectral types of a number of stars are measured.  Stars with the same spectral type are expected to have the same intrinsic colors. The reddening of stars, as given by the difference between the observed colors of stars and the intrinsic colors of stars of the same spectral type, is attributed to dust.  By building up a large sample of stars, we can map the distribution of dust in the Milky Way. 

The pair method has a long history in astronomy, dating back to the work of \cite{Trumpler_1930}.  A common approach is to directly compare reddened stars with lightly reddened or unreddened stars to determine extinction \citep[e.g.,][]{Trumpler_1930, Stecher:1965, Bless&Savage:1970, Massa_Savage_Fitzpatrick:1983}. A related approach involves training a mathematical model on unreddened stars and using it to predict instrinsic colors of similar reddened stars so that the difference between the observed and predicted colors provides extinction due to dust \citep[e.g.,][]{Schlafly:2011, Yuan:2013, Schlafly:2016}.  Related approaches also work for galaxies \citep[e.g.,][]{Peek&Graves:2010}.  Another common approach is to obtain a prediction for the intrinsic color using a stellar atmosphere model (calibrated to observational stellar data) and difference it with the observed colors of reddened stars to obtain extinction (\citealt{Fitzpatrick_Massa:2007, Wang:2023, Zhou_2025}). 

Extinction depends on wavelength and this dependence is described by the extinction curve, which is often parameterized by a single number: the total-to-selective extinction ratio, $R(V)$ \citep{CCM:1989}. The shape of the extinction curve probes the properties of dust grains.   Variation in $R(V)$ is most commonly attributed to the grain size distribution: high $R(V)$ sightlines have flatter optical extinction curves, indicating a grain population weighted toward larger sizes \citep{Weingartner:2001, Draine:2003}. Other proposed drivers include the growth of ice mantles and grain coagulation in dense environments
\citep{Whittet:1988, Ysard:2013}, and the processing of grains by
ultraviolet photons or differences in grain composition and chemistry
\citep{Jones:2013, Mulas:2013}.

The mapping of $R(V)$ over large areas of the sky has been done by two principal approaches. The first approach involves measuring reddenings in several colors and converting their ratio into $R(V)$ using an assumed family of extinction curves. Here, the stellar type is determined beforehand, so the extinction measurement and the stellar type determination are separate steps. \cite{Schlafly:2016} took this approach, combining APOGEE spectroscopy with Pan-STARRS1, 2MASS, and \wise photometry to map $R(V)$ over a low Galactic latitude footprint. Similarly, \citet{Zhang:2023} combined LAMOST spectroscopy with \gaia, Pan-STARRS1, SDSS, 2MASS, and \wise photometry to map $R(V)$ over the northern sky, with reliable measurements concentrated at low and intermediate Galactic latitudes. The second approach involves fitting the shape of the extinction curve simultaneously with the stellar parameters. \cite{Zhang&Green:2024} applied this to \gaia XP spectrophotometry and mapped $R(V)$ over the whole sky, including the high Galactic latitude regime where extragalactic surveys operate. However, at the low resolution ($R < 100$) of those data, stellar type and extinction curve are covariant, complicating the analysis. In this work, we instead use the first approach, using the largest spectroscopic data set ever employed at high Galactic latitudes.

In this work, we use the Dark Energy Spectroscopic Instrument (DESI) \citep{Levi:2013}, which has currently observed more than 40 million galaxies and 10 million stars. We use these stars in a pair-method-like approach to map dust. Recent work by \cite{Zhou_2025}, henceforth Z25, also computed reddening based on DESI spectroscopy and photometry, and published a dust map that improves dramatically on the traditional SFD thermal dust map \citep{SFD_1998}. They show that on large scales, 80\% of the observed fluctuations in the density of a particular tracer of the large scale structure is explained by errors in the SFD dust map; errors in dust maps are the dominant source of noise in today’s cosmological measurements at large scales.  Adoption of a dust map directly based on optical extinction measurements eliminates these errors.

The work of Z25 has focused on delivering the extinction corrections necessary to mitigate extinction systematics in the DESI cosmology analysis.  Our work differs from theirs in two ways. First, Z25 fit theoretical stellar atmosphere models to observed spectra for predicting intrinsic colors, while we use empirically-derived intrinsic colors from low-reddening calibration stars. Second, Z25 focus on reporting improvements over SFD maps, while we extend our results to the shape of the extinction curve and its variation over the DESI footprint. 

In this work, we develop a machine-learning based approach to the ``pair method,'' and use it to measure the extinction toward stars with DESI spectroscopy. We use XGBoost \citep{xgboost:2016} to learn the relationship between the temperatures, metallicities, and surface gravities of stars as measured by DESI, and their intrinsic colors as observed by the DESI Legacy Imaging and \wise surveys. This intrinsic-color relation is trained on a low-reddening calibration data set where dust is expected to be unimportant, and then applied to the entire DESI sample.  By differencing the predicted intrinsic colors with the observed colors, we measured the extinction curve to 3.3M stars, with an accuracy of 20 mmag in the $g-r$ and $r-z$ colors.  We then use those results to map extinction over the DESI footprint, similar to Z25 but with an alternative approach. 

We then develop a forward model to map these derived reddenings to $R(V)$ and create $R(V)$ maps over the regions of the DESI footprint with significant dust clouds. We study the variation in $R(V)$ and explore the potential causes for it.

This paper is structured as follows. In \textsection\ref{sec:data} we discuss the sources of the data used and the quality cuts placed on it for selecting our final sample of stars. We lay out the machine learning model for predicting intrinsic colors of stars in \textsection\ref{sec:extinctions_reddening}, and use it to compute the reddenings of stars. In \textsection\ref{sec:rv_and_variation}, we compute $R(V)$ and study its variation, and in \textsection\ref{sec:discussion}, we attempt to correlate the variation in $R(V)$ with that of other probes of the interstellar medium and discuss the implications of our work on far-IR-based dereddening. We explore the sources of systematic uncertainty in our results in \textsection\ref{sec:sys}. In \textsection\ref{sec:data_availability}, we share the data product for this work. Finally, in \textsection\ref{sec:conclusion}, we provide the overall conclusions.

\needspace{4\baselineskip}
\section{Data}
\label{sec:data}

\subsection{Source Surveys and Instruments}
\label{subsec:surveys}
We  the imaging data provided by Legacy Survey DR9 imaging \citep{LS.Overview.Dey.2019} for the photometry in $g$, $r$ and $z$ bands. The Legacy Survey Imaging consists of two photometric systems: North and South, separated at a declination of 32.375$^\circ$. For the northern imaging, the $g$ and $r$ band data comes from the Beijing-Arizona Sky Survey \citep[BASS;][]{Zou_2017} and the $z$ band data comes from the Mayall $z$-band Legacy Survey \citep[MzLS;][]{LS.Overview.Dey.2019}, both conducted on Kitt Peak in Arizona. The Southern Imaging comes from a combination of the Dark Energy Camera Legacy Survey \citep[DECaLS;][]{LS.Overview.Dey.2019}, the Dark Energy Survey \citep[DES;][]{10.1093/mnras/stw641}, and additional DECam programs. The mid-infrared imaging comes from the Wide-Field Infrared Survey Explorer \citep[\wise;][]{wright2010}, where we use the $W1$ and $W2$ bands. The photometry for \wise is based on the methods presented in \cite{Lang:2016}.

For spectroscopic analysis, we utilize the data observed and processed \citep{Guy:2023} by The Dark Energy Spectroscopic Instrument \citep[DESI;][]{DESI2022.KP1.Instr}, which is a multi-object spectrograph on the 4 m Mayall telescope at Kitt Peak National Observatory \citep{Miller:2024}. It is composed of robotic fibers that can capture spectra of almost 5000 astronomical objects simultaneously over a $\sim$3$^\circ$ field (\citealt{DESI2016b.Instr, FiberSystem.Poppett.2024}). The instrument is conducting a survey covering a footprint of 17,000 $\deg^2$ of the sky, and began publicly releasing results in 2023 \citep{desiedr, desicollaboration2026datarelease1dark, Schlafly:2023}, together with associated cosmological constraints \citep{DESI2024.VII.KP7B, DESI:2025PhRvD.112h3515A}.

\subsection{Data Preprocessing}
\label{subsec:preprocessing}

\begin{table}[ht!]
\centering
\begin{tabular}{l >{\raggedright\arraybackslash}p{0.5\columnwidth}}
\hline
Catalog & Role \\
\hline
zall-pix-loa & Base DESI catalog \\
rvpix-loa & Used for spectroscopic stellar parameters such as metallicity \\
tractorphot & Used for detailed photometry flags such as fractional flux\\
\citealt{Zhou:2023b} & Used for zero-point corrections to the DESI imaging using \gaia \\
\hline
\end{tabular}
\caption{Datasets used for determining the reddenings in this work.  We use products associated with the \texttt{loa} processing that will eventually become DESI DR2, together with photometric corrections to the DESI DR9 imaging.}
\label{table:datasets}
\end{table}

The datasets used in this work are listed in Table \ref{table:datasets}. We use the DESI \texttt{loa} internal data reduction, which will eventually become DR2. Our base catalog is \texttt{zall-pix-loa}, which includes target spectra from \texttt{loa} across all surveys and programs, together with flags to identify the best recommended redshift for duplicate observations of the same target. It consists of $\sim$64.1M targets. We merge this with two Value Added Catalogs. The first is the DESI Stellar Catalogue, from which we exclusively use the \textit{RVSpecFit} (\texttt{rvpix-loa}) processing \citep{Koposov:2024}. The catalog provides stellar parameters and radial velocities measured by the RVS pipeline (\citealt{Koposov:2011, Koposov:2019}). The version of this pipeline applied to \texttt{loa} differs only in minor respects from the version applied to DESI DR1 \citep{Koposov:2025}. The RVS pipeline fits each spectrum while marginalizing over the continuum shape, so the resulting stellar parameters are set by absorption-line features rather than by the broadband spectral shape and are unaffected by dust extinction. The second Value Added Catalog we use is the \texttt{tractorphot} photometric catalog \citep{Dey:2019}, which provides detailed photometry, including flags such as fractional flux. Merging with these two catalogs eliminates most galaxies, reducing the dataset to $\sim$12.4 M targets.

\begin{table*}[htb]
    \caption{Cuts placed on the data}
    \centering
    \begin{tabular}{@{}>{\raggedright\arraybackslash}p{0.5\textwidth} 
                        >{\raggedright\arraybackslash}p{0.45\textwidth}@{}}
        \toprule
        \textbf{Cut} & \textbf{Comment} \\
        \midrule

        \multicolumn{2}{c}{\text{Sample and target selection cuts}} \\

        SPECTYPE = STAR 
        & Select only stars, not galaxies and quasars \\

        DESI\_TARGET $\in$ \{STD\_FAINT, STD\_BRIGHT, MWS\_ANY\} 
        & Select DESI targets in the Milky Way and standard star programs to ensure a uniform, well-characterized stellar photometric selection. \\

        PROGRAM $\neq$ backup & Exclude backup program targets \\

        MWS\_TARGET $\notin$ \{MWS\_WD, MWS\_BHB, MWS\_NEARBY\} 
        & Exclude white dwarfs and blue horizontal branch stars due to their unique spectral features and colors, as well as nearby stars that may be in the foreground of the dust column. \\

        \midrule
        \multicolumn{2}{c}{\text{Spectroscopic quality and parameter cuts}} \\

        SUCCESS = True 
        & Select stars with no warning flags in the stellar parameter fitting from \citet{Koposov:2025}. \\

        $\teff \in (4000, 7000)\,\text{K}$, 
        $\feh \in (-3.0, 0.3)$, 
        $\logg \in (2.5, 5.7)$ 
        & Retain only stars with typical temperatures, metallicities, and gravities. \\

        \midrule
        \multicolumn{2}{c}{\text{Photometric quality cuts}} \\

        $\text{FLUX}\_i \ge 5\sigma(i)\; 
        \forall i \in \{\text{g, r, z, W1}\}$ 
        & Limit flux to 5-sigma values for all bands other than W2 (which is shallower and unused in our $R(V)$ measurements). \\

        $\text{FLUX\_W2} > 0$ 
        & Ensure W2 flux is at least positive so magnitudes are well defined. \\

        $\text{FRACFLUX}\_i < 0.1 \ 
        \forall i \in \{\text{g, r, z}\}$ 
        and $\text{FRACFLUX}\_\mathrm{W1} < 0.8$ 
        & Reject objects whose photometry is contaminated by the flux of nearby, overlapping sources. Use a more lenient limit for $W1$ due to the comparatively broader \wise point spread function. \\

        $\text{FRACMASKED}\_i < 0.25 \ 
        \forall i \in \{\text{g, r, z}\}$ 
        & Reject objects that fall on bad pixels, cosmic rays, or other defects that have compromised the structural fit. \\

        $\text{SATUR}\_i \ 
        \forall i \in \{\text{g, r, z}\}$ and WISEM1 bits not set 
        & Ensure that bands $g$, $r$, $z$, and $W1$ are unsaturated. \\

        \bottomrule
    \end{tabular}
    \label{table:cuts}
\end{table*}

\begin{figure*}[t]
    \centering
    \gridline{\fig{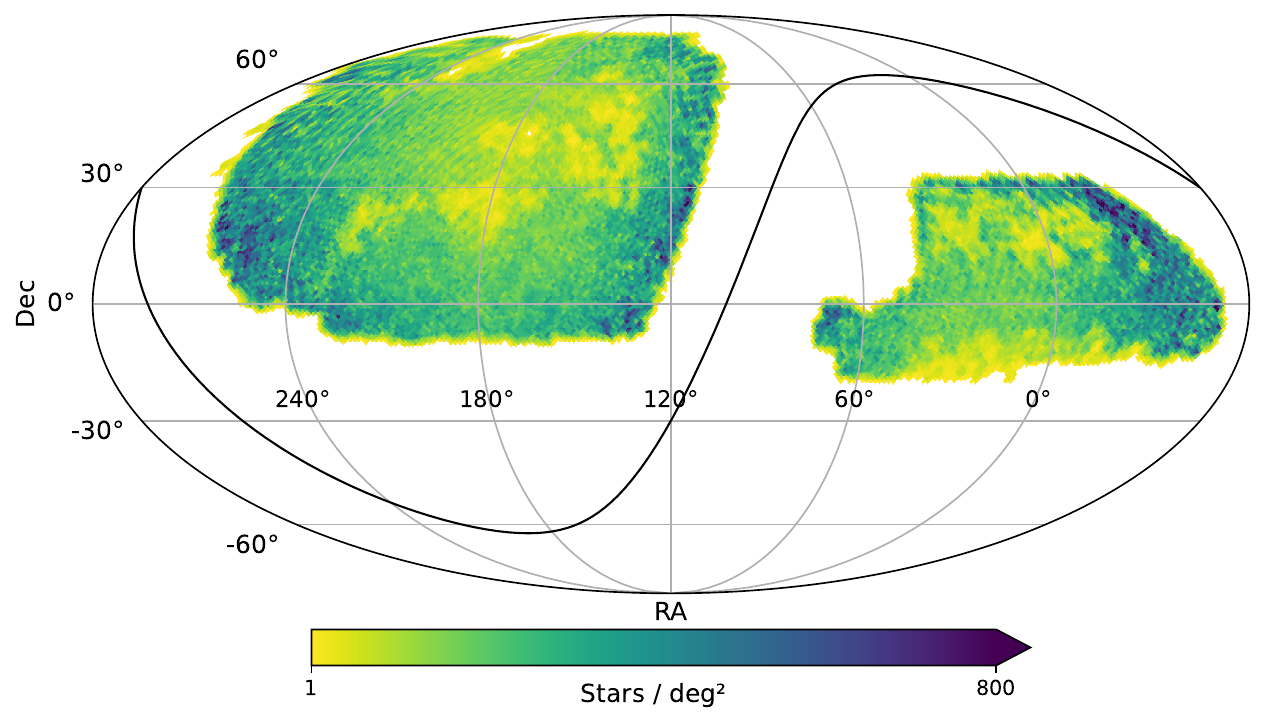}{0.78\textwidth}{(a)}}
    
    \gridline{\fig{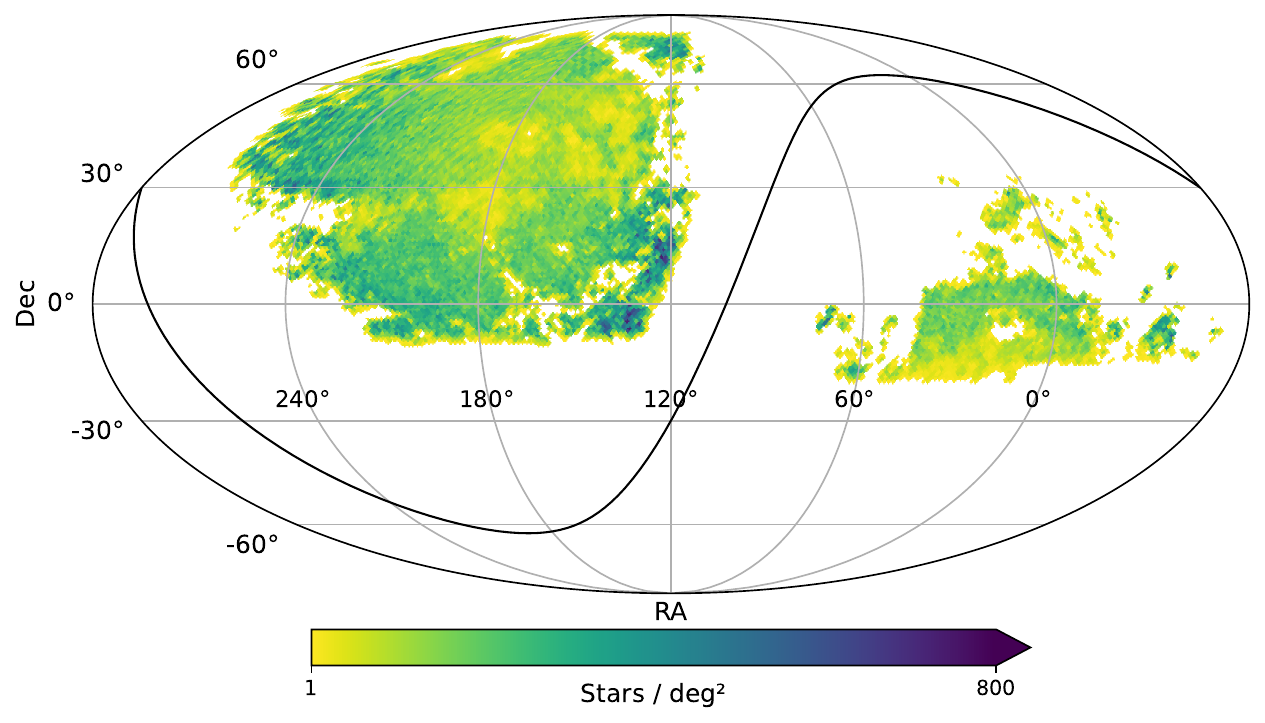}{0.78\textwidth}{(b)}}
    
    \caption{Comparison of the full selected DESI footprint of stars and calibration stars used in this work, shown in celestial coordinates. 
    \textbf{(a)} Set of 3.3M stars over the DESI footprint selected for analysis in this work. 
    \textbf{(b)} Set of 1.5M low-reddening ($\ebv < 0.04$) stars called \textit{calibration stars} selected for training the intrinsic color model.}
    \label{fig:all_calibration_star_comparison}
\end{figure*}

We then impose the quality cuts listed in Table \ref{table:cuts} on these 12.4 M targets. Broadly, these cuts aim to select stars rather than galaxies; to select ``normal'' stars, avoiding special classes of nearby stars or stars with more complicated target selections; and to select stars with good photometry and spectroscopy. Of the 12.4M targets,  $\sim$51$\%$ do not meet the DESI and MWS target selection criteria (mostly BACKUP program targets), $\sim$12$\%$ lie outside the stellar parameter limits imposed, and $\sim$9$\%$ have poor photometric quality. The quality cuts result in 3.3M stars (see Figure~\ref{fig:all_calibration_star_comparison}a). Of these, approximately half (1.5M) of the stars (see Figure~\ref{fig:all_calibration_star_comparison}b) are in the low-reddening region, with $\ebv < 0.04$.

To mitigate systematic errors arising from spatially-varying photometric zero-points in the Legacy Surveys, we then calibrate our $g$, $r$, and $z$-band magnitudes to the \gaia photometric system using the maps of \cite{Zhou:2023a, Zhou:2023b}. To do so, we take the maps of median magnitude difference between Legacy Survey and \gaia at a HEALPix \citep{HEALPIX:2005} \texttt{nside} of 128, enforce a minimum of 4 objects per pixel, and apply the zero point correction from the map to the observed magnitudes by interpolating at the location of each star.

\section{Extinctions and reddenings}
\label{sec:extinctions_reddening}
Our goal is to study the extinction curve along different lines of sight at the high Galactic latitudes observed by DESI. The extinction curve is often parameterized by the extinction ratio $R(V)$\citep{CCM:1989}, defined as the ratio of the total to selective extinction
\begin{equation}
    R(V) = \frac{A(V)}{E(B-V)} \, ,
\label{eq:rv}
\end{equation}
where $A(V)$ is the extinction in the $V$ band and 
$E(B-V)$ is the reddening in $B-V$.

As the DESI Legacy Imaging Surveys do not provide data for the $B$ and $V$ bands, we cannot directly get $A(V)$ and $E(B-V)$ to compute $R(V)$ per Equation~\eqref{eq:rv}. So, in this section, we develop a method to derive extinction and reddening for the bands available for DESI sources (\textsection\ref{subsec:colormodel}) and validate it (\textsection\ref{subsec:colormodel_validation}). We then use the measurements to map reddening over the DESI footprint (\textsection\ref{subsec:reddenings_eval_stars}) and compare with results from the literature (\textsection\ref{subsec:comparison_with_sfd_rongpu}) and with results derived in other filters (\textsection\ref{subsec:egr_erz_egw1}).  In subsequent sections, we then use the resulting maps to study $R(V)$.

\subsection{Intrinsic Color Modeling}
\label{subsec:colormodel}
\begin{figure*}[htb!]
    \centering
    \includegraphics[width=\textwidth]{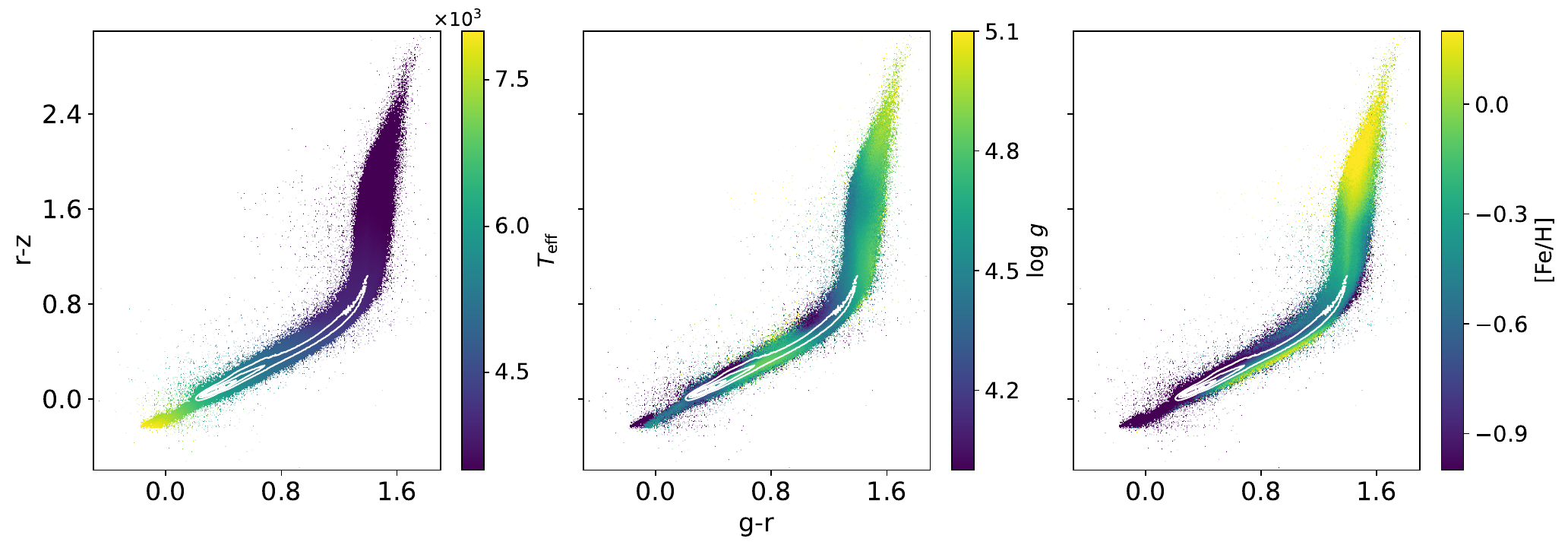}
    \caption{Color-color diagrams for 4.3M stars observed by DESI and dereddened using SFD. The diagrams are colored by the stars' temperatures, metallicities, and gravities. Note that we use a broader cut on spectroscopic parameters than in Table~\ref{table:cuts} in order to show more global shape of the stellar locus.}
    \label{fig:color_color}
\end{figure*}

To compute reddenings, we require the observed and intrinsic colors of stars. While the observed colors are available in the DESI Legacy Imaging Surveys, computing intrinsic colors is more challenging.  In order to derive these, we rely on the fact that the spectral energy distribution of a star can largely be determined from its temperature, metallicity, and surface gravity. We show the broadband colors of stars as a function of their spectroscopic parameters \citep{Koposov:2024, Koposov:2025} in Figure~\ref{fig:color_color}.  As expected, the colors of stars vary smoothly with their spectroscopic parameters. Roughly speaking, a star's temperature determines its location along the stellar locus in Figure~\ref{fig:color_color}, while its metallicity determines its location perpendicular to the locus for main sequence stars (see also Figure 12 from \citealt{Li_Koposov:2019}).  The star's surface gravity provides some additional information.  By combining these three spectroscopic parameters, we expect to be able to predict the intrinsic colors of a star closely.

We model the intrinsic color, $c_\mathrm{int}$, of a star as a function of its spectroscopic parameters using
\begin{equation}
    c^\mathrm{mod}_{\mathrm{int}} = h(\teff, \feh, \logg)
\label{eq:colormodel}
\end{equation}
where the three spectoscopic parameters chosen are the effective temperature (\teff), metallicity (\feh), and surface gravity (\logg), and $c$ represents the color formed by two bands. We consider a specific set of colors in this paper and perform the analysis on each color independently.

To train the model, we construct intrinsic colors by dereddening observed colors using SFD over a low-reddening ($\ebv <0.04$) portion of the footprint:
\begin{equation}
    c_{\mathrm{int}} = c_\mathrm{obs} - E(c)_\mathrm{SFD}
\label{eq:intrinsiccolor}
\end{equation}
where $c_\mathrm{obs}$ is the observed color and $E(c)_\mathrm{SFD}$ is obtained using
\begin{equation}
    E(c)_\mathrm{SFD} = (R_{x}-R_{y}) \cdot E(B-V)_\mathrm{SFD}
\label{eq:ebvtoexy}
\end{equation}
Equation~\eqref{eq:ebvtoexy} converts $\ebv$, the reddening in $B-V$ obtained from SFD, to $E(c)_\mathrm{SFD}$, the reddening in color $c = x-y$, assuming an average extinction curve. We obtain $R_x$ from \cite{Schlafly:2011}\footnote{\url{https://www.legacysurvey.org/dr10/catalogs/\#galactic-extinction-coefficients}}.
We compute reddenings for colors $g-r$, $r-z$, $z-W1$, $W1-W2$, $g-W1$, and $g-W2$, and the coefficients used for converting $\ebv$ to those reddenings are 1.049, 0.954, 1.027, 0.071, 3.03, and 3.101 respectively.

\begin{figure*}
    \centering
    \includegraphics[width=\textwidth]{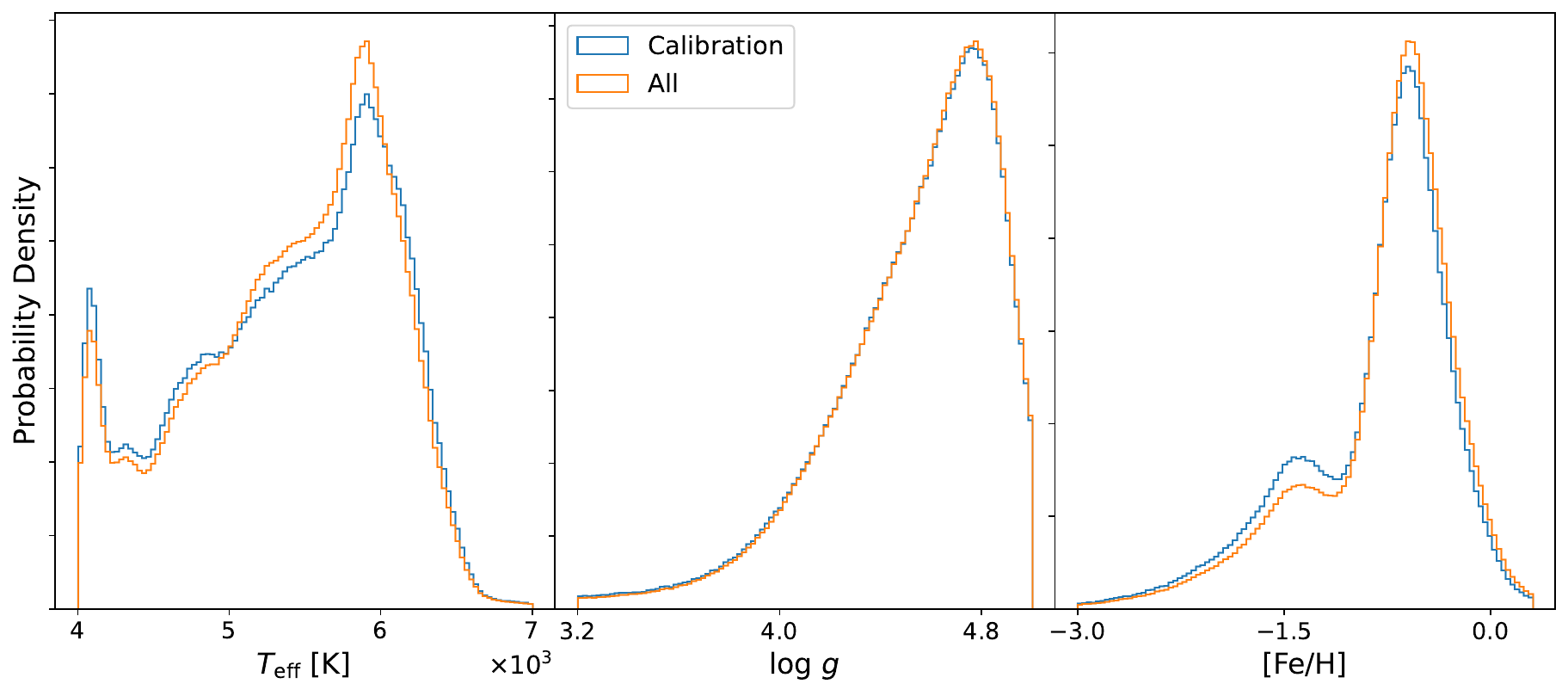}
    \caption{The distribution of $\teff$, $\logg$, and $\feh$ for calibration stars and the full set (including calibration) of stars in our analysis. The plots show that the calibration stars selected for training are reasonably matched to the distribution of all stars observed by DESI.}
    \label{fig:calib_vs_all_hist}
\end{figure*}

To train the model, we use 1.5M low-reddening ($\ebv < 0.04$) stars (see Figure~\ref{fig:all_calibration_star_comparison}b) with DESI spectroscopy. We call these \textit{calibration stars}. We wish to apply a fit derived from these calibration stars to our full DESI sample of stars.  We therefore want these calibration stars to be representative of the full sample.  Figure~\ref{fig:calib_vs_all_hist} compares the distribution of stellar parameters of the calibration stars with that of all the stars in our dataset.  The distributions of temperature, metallicity, and surface gravity are very similar between the calibration set and the full set of stars. 

To train the model in Equation~\eqref{eq:colormodel}, we use an XGBoost regressor with the intrinsic color set as the target and spectroscopic parameters as the features. All features are scaled to have zero mean and unit variance. The model uses 1000 estimators, a maximum depth of 6, a subsampling of 80\%, a learning rate of 0.03, and a pseudo-Huber loss function. The loss function has a Huber $\delta$ of 1, so it behaves as $L2$ (mean squared error) for small residuals and transitions to $L1$ (mean absolute error) for residuals larger than 1, limiting the influence of outliers. We use the standard train/test split of 80/20 on the calibration stars. We perform this analysis for the colors $g-r$, $r-z$, $z-W1$, $W1-W2$, $g-W1$, and $g-W2$.

The northern and southern portions of the DESI Legacy Imaging Surveys were observed with different telescopes and filters---DECam ($g$, $r$, $z$) in the South, and BASS ($g$, $r$) and MzLS ($z$) in the North. These differences imply that the intrinsic stellar locus---the relationship between stellar parameters (\teff, \feh, \logg) and observed color---is slightly different for each footprint. A single global model trained on the combined dataset would ignore these variations, conflating physical stellar properties with instrumental systematics and inflating the scatter in our predictions. To accurately capture these instrument-specific baselines, for each color, we train two independent XGBoost regressors: one exclusively on calibration stars with the northern photometric system i.e. $\mathrm{PHOTSYS}=N$, and another on calibration stars with $\mathrm{PHOTSYS}=S$. By learning separate mapping functions, $h_N$ and $h_S$ (Equation~\ref{eq:colormodel}), we ensure that our intrinsic color predictions, $c^\mathrm{mod}_{\mathrm{int}}$, are tailored to the specific bands and calibration of each survey region. 

An important feature of this approach to modeling the intrinsic colors of stars is that we are largely insensitive to biases in the spectroscopic temperature, metallicity, and surface gravity measurements.  For example, our results would be unaffected if all of the temperatures were off by a scale factor or constant offset.  We rely only on the existence of a mapping from the spectroscopic parameters to intrinsic stellar colors, which the machine-learning algorithm learns.

\subsection{Intrinsic Color Model Validation}
\label{subsec:colormodel_validation}

To evaluate the predictive performance of our model, we compare the intrinsic color, $c_\mathrm{int}$, given by Equation~\eqref{eq:intrinsiccolor}, with the  intrinsic color predicted by the model, $c^\mathrm{mod}_\mathrm{int}$, given by Equation~\eqref{eq:colormodel}. We quantify the fraction of variance captured by the model using the coefficient of determination
\begin{equation}
R^2 = 1 - \frac{ \sum_i \left(c_{{\rm int},i} - c_{{\rm int},i}^{\rm mod}\right)^2}
{\sum_i \left(c_{{\rm int},i} - \overline{c}_{\rm int}\right)^2}
\label{eq:r2_score}
\end{equation}
where $i$ indexes the stars and $\overline{c}_{\rm int}$ is the mean intrinsic color of the star sample.

Here, the sums are taken over the stars in the corresponding training or test sample. We also define the model residual for each color $c$ as
\begin{equation}
    \Delta_{c} = c_\mathrm{int} - c^\mathrm{mod}_\mathrm{int}
\label{eq:modelresidual}
\end{equation}
and characterize the residual scatter by computing the normalized median absolute deviation, $\sigma_{\rm NMAD}$, of $\Delta_{c}$.

\begin{table}
\centering
\caption{Model residuals}
\label{tab:model_residuals}
\begin{tabular}{cccccc}
\hline
\multirow{2}{*}{Color $c$} & 
\multirow{2}{*}{PHOTSYS} & 
\multicolumn{2}{c}{$R^2$} & 
\multicolumn{2}{c}{$\sigma_{\rm NMAD}(\Delta_{c})$} \\
\cline{3-6}
 & & Train & Test & Train & Test \\
\hline

\multirow{2}{*}{$g-r$}
& N & 0.994 & 0.993 & 18 & 18 \\
& S & 0.994 & 0.994 & 17 & 18 \\
\hline

\multirow{2}{*}{$r-z$}
& N & 0.984 & 0.984 & 22 & 22 \\
& S & 0.986 & 0.985 & 18 & 19 \\
\hline

\multirow{2}{*}{$z-W1$}
& N & 0.883 & 0.881 & 57 & 57 \\
& S & 0.891 & 0.890 & 53 & 53 \\
\hline

\multirow{2}{*}{$W1-W2$}
& N & 0.030 & 0.004 & 103 & 104 \\
& S & 0.023 & 0.009 & 104 & 105 \\
\hline

\multirow{2}{*}{$g-W1$}
& N & 0.978 & 0.977 & 81 & 82 \\
& S & 0.978 & 0.978 & 77 & 77 \\
\hline

\multirow{2}{*}{$g-W2$}
& N & 0.905 & 0.904 & 145 & 146 \\
& S & 0.873 & 0.866 & 145 & 146 \\
\hline

\end{tabular}

\vspace{1ex} 
    \parbox{\columnwidth}{\footnotesize 
        \textbf{Note.} $R^2$ is the coefficient of determination and
$\sigma_{\rm NMAD}$ is the normalized median absolute deviation (in mmag) of
the raw model residuals, $\Delta_{c}$. The training and test samples
contain 80\% and 20\% of the calibration stars for a given PHOTSYS, respectively.
}
\end{table}

Table~\ref{tab:model_residuals} shows $R^2$ and $\sigma_{\rm NMAD}$ for calibration stars for each color and photometric system. We find $R^2 > 0.85$ for both the training and test samples for all colors except $W1-W2$, demonstrating that the models account for most of the variance in the intrinsic color. The performance on the training and test samples is also very similar; overfitting is not a problem.

For $W1-W2$, the $R^2$ values are close to 0, indicating that the model has little predictive power for this color.  This is because $W1-W2$ probes the Rayleigh-Jeans tail of stellar spectral energy distribution for the relatively hot stars considered here; the intrinsic colors are close to constant, and most of the observed variation in color is driven by photometric uncertainties.

Taken together, the close agreement between the training and test metrics for the colors indicates that the models generalize well to unseen calibration stars and show no significant evidence of overfitting. Furthermore, Figure~\ref{fig:calib_vs_all_hist} shows that the calibration set is broadly representative of the full selected DESI footprint in the spectroscopic parameters used by the model, giving us confidence that the models should perform similarly when applied to the full set of stars.

The raw residual, $\Delta_{c}$, however, contains contributions not only from the intrinsic model uncertainty but also from large-scale systematic uncertainties in SFD and the Legacy Survey photometric calibration. To isolate the model's precision from these effects, we calculate a locally corrected residual. First, we compute the residuals over the entire set of calibration stars using Equation~\eqref{eq:modelresidual}. Next, we apply a smoothing kernel (FWHM $= 1^\circ$) to the raw residual map to estimate the local bias, $\bar{\Delta}_c$, which is then subtracted from the individual stellar residuals, $\Delta_c$, to produce the small-scale residual fluctuation, $\delta_c$:
\begin{equation}
    \delta_c = \Delta_c - \bar{\Delta}_c
\label{eq:small_scale_model_residual}
\end{equation}
This high-pass filtering removes regional calibration drifts and highlights the local star-to-star precision of the model. Table \ref{tab:locally_corrected_residuals} shows these results for calibration stars for all colors. Figure~\ref{fig:rongpu_maps_comparison}(a) shows a map of $\Delta_c$ for $g-r$.

\begin{figure*}[t!]
    \centering
    \includegraphics[width=\textwidth]{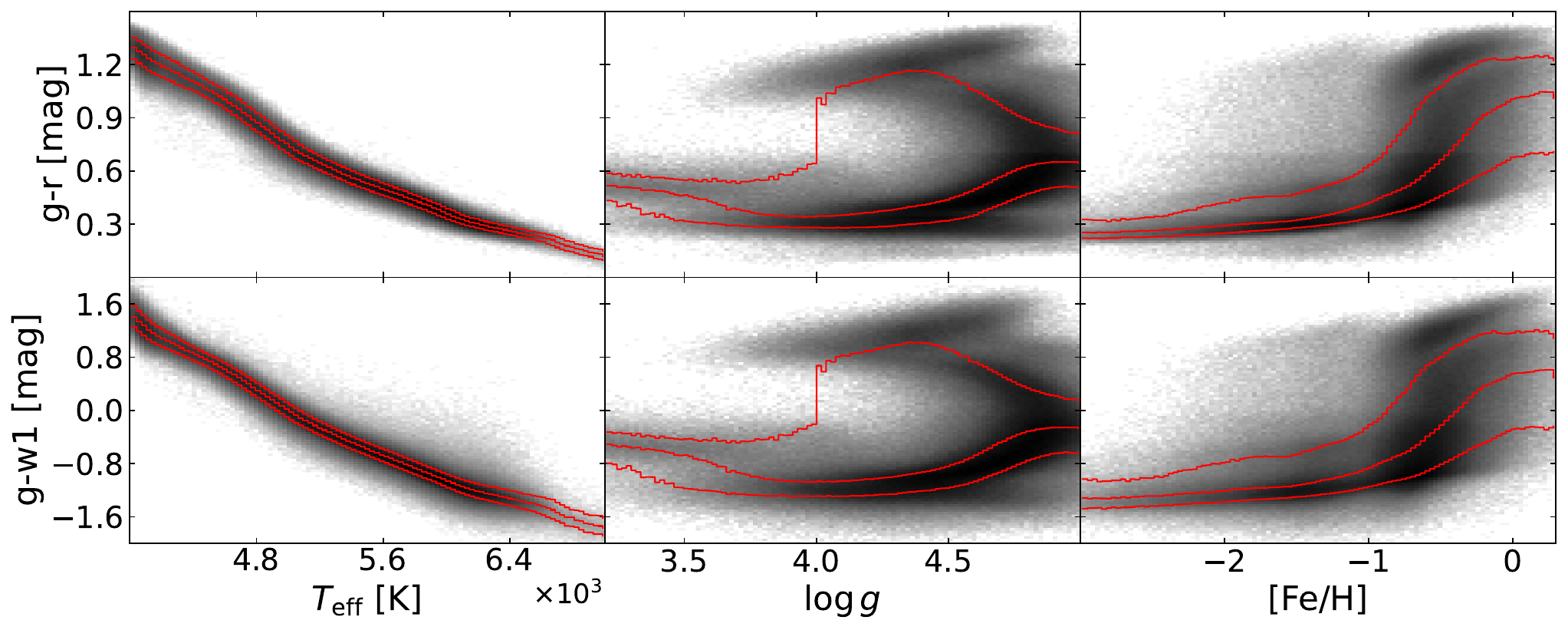}
    \caption{The variation in $g-r$ and $g-W1$ against different spectroscopic parameters for calibration stars. The red lines represent the 16th, 50th (median) and 84th percentiles. Significant variation of the colors against these parameters, especially temperature, demonstrates why these parameters are good predictors of the colors.}
    \label{fig:colors_against_stellar_params}
\end{figure*}

\begin{figure*}[t!]
    \centering
    \includegraphics[width=\textwidth]{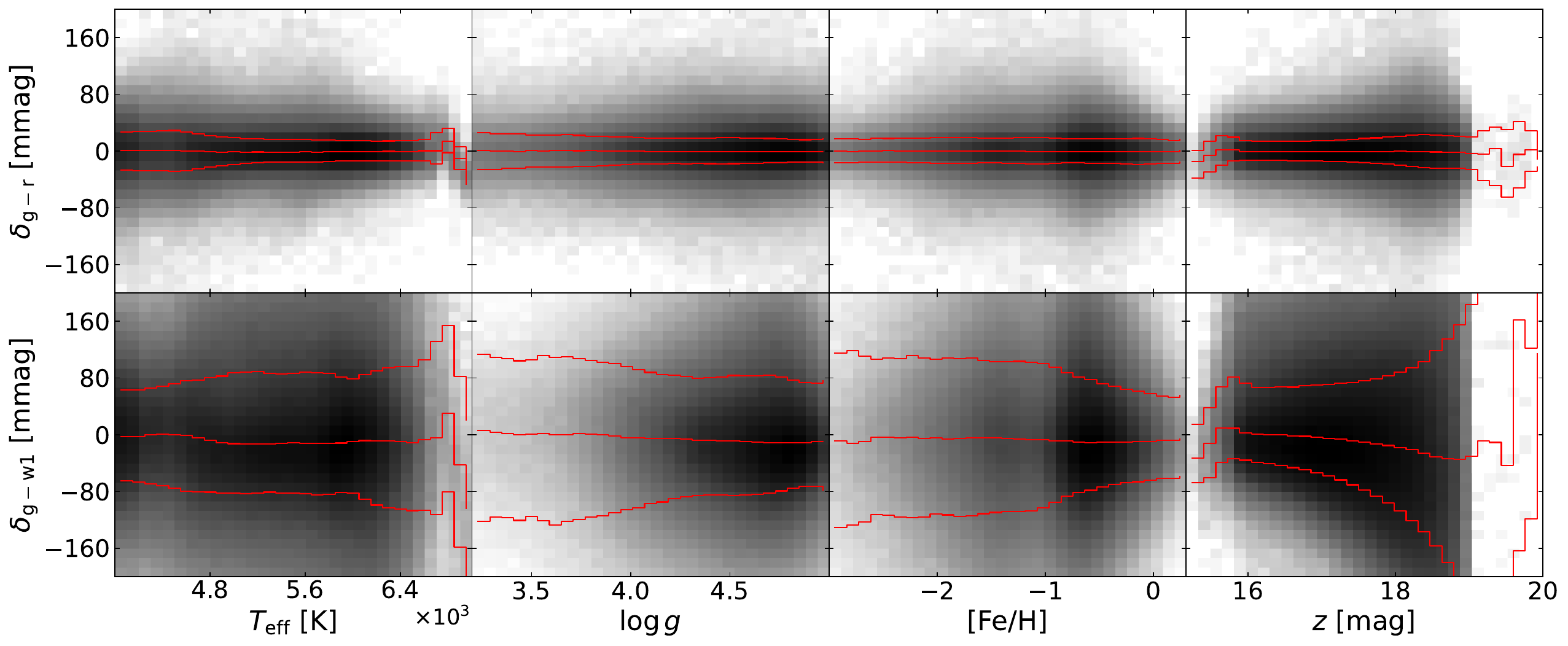}
    \caption{The locally corrected residuals $\delta_c$ (Equation~\ref{eq:small_scale_model_residual}) for $g-r$ and $g-W1$, shown against spectroscopic parameters and $z$ magnitude for calibration stars. The red lines represent the 16th, 50th (median) and 84th percentiles. Unsurprisingly, $g-W1$ shows larger scatter than $g-r$ due to the lower sensitivity of \wise relative to the DESI Legacy Imaging Surveys, and the greater temperature sensitivity of that color. The higher scatter in $g-W1$ towards the faint end of $z$ comes from photometric uncertainties in \wise.}
    \label{fig:residualplots}
\end{figure*}

For studying $R(V)$ (\textsection\ref{sec:rv_and_variation}), we only use the $g-r$ and $g-W1$ colors; we focus on these two colors for the remainder of this work.  Figure~\ref{fig:colors_against_stellar_params} shows the variation of these colors with temperature, metallicity, and surface gravity. Figure~\ref{fig:residualplots} shows the model residuals $\delta_c$ against these same parameters and against $z$ magnitude. The variation in color is captured well by our model, resulting in flat residuals with small scatter. The scatter ranges from 17~mmag in $g-r$ to 60--150~mmag in $g-W1$, depending on the brightness of the source.  The larger scatter in $g-W1$ is due to the greater temperature and reddening sensitivity of $g-W1$ than $g-r$ and the larger photometric uncertainties in the \wise bands relative to the optical bands, which we investigate in more detail in \textsection\ref{subsec:uncfloor}.  The photometric uncertainties in \wise become dominant near the faint limit in $z$, which is why we see the scatter in $g-W1$ rise rapidly for $z > 17.5$~mag.  Median residuals are slightly offset from zero due to a modest skew in the distribution of residuals, but this offset is only $\sim 0.1\sigma_{\rm{NMAD}}$ for all colors except $W1-W2$ (Table~\ref{tab:locally_corrected_residuals}).

\begin{table}[ht!]
    \centering
    \caption{Locally corrected model residuals}
    \label{tab:locally_corrected_residuals}
    
    \begin{tabular*}{\columnwidth}{@{\extracolsep{\fill}}ccc} 
        \toprule
        \toprule
        Color $c$ & $\sigma_{\mathrm{NMAD}}(\delta_c)$ & Med($\delta_c$) \\
        \midrule
        $g-r$   & 17 & -1 \\
        $r-z$   & 19 & -3 \\
        $z-W1$  & 54 & -5 \\
        $W1-W2$ & 105 & 24 \\
        $g-W1$  & 77 & -9 \\
        $g-W2$  & 146 & 16\\
        \bottomrule
    \end{tabular*}
    
    \vspace{1ex} 
    \parbox{\columnwidth}{\footnotesize 
        \textbf{Note.} Summary statistics describing our model's precision. $\delta_c$ is computed using Equation~\eqref{eq:small_scale_model_residual} and represents the local scatter per star relative to the neighborhood average for calibration stars. $\sigma_{\rm{NMAD}}$ and Med represent the normalized median absolute deviation and median respectively, both in mmag.
    }
\end{table}

\subsection{Computing Reddenings over the Full Footprint}
\label{subsec:reddenings_eval_stars}
We apply the model developed in \textsection\ref{subsec:colormodel} to the set of 3.3M stars over the full footprint (see Figure~\ref{fig:all_calibration_star_comparison}a) to obtain their intrinsic colors
$c^\mathrm{mod}_\mathrm{int}$. The reddening in color $c$, $E(c)$, is obtained as the difference between the observed and predicted intrinsic color: 
\begin{equation}
    E(c) = c_\mathrm{obs} - c^\mathrm{mod}_\mathrm{int}
\label{eq:reddening}
\end{equation}

We then construct reddening maps in different colors by binning stars into different pixels and averaging the resulting reddening measurements in each pixel.  The difference between the DESI Legacy Imaging Survey northern and southern photometric systems, as described in \textsection\ref{subsec:colormodel}, introduces a discontinuity in reddening at the survey boundary,  $\delta = 32.375^\circ$. To address this, we compute the difference between the median reddening residual after subtracting SFD in a band $10^\circ$ south and north of $\delta = 32.375^\circ$.  We then apply this offset to the northern footprint to remove the discontinuity.
The correction applied for $g-r$ and $g-W1$ is 4.2 and 11.4~mmag respectively. Figure~\ref{fig:egr_comparison}a shows the final map of the derived reddening in the $g-r$ color at a resolution of $55'$ (HEALPix $\texttt{nside} =64$). 

\subsection{Comparison with Reddening Maps in the Lifterature}
\label{subsec:comparison_with_sfd_rongpu}
\begin{figure*}[t]
    \centering
    \gridline{\fig{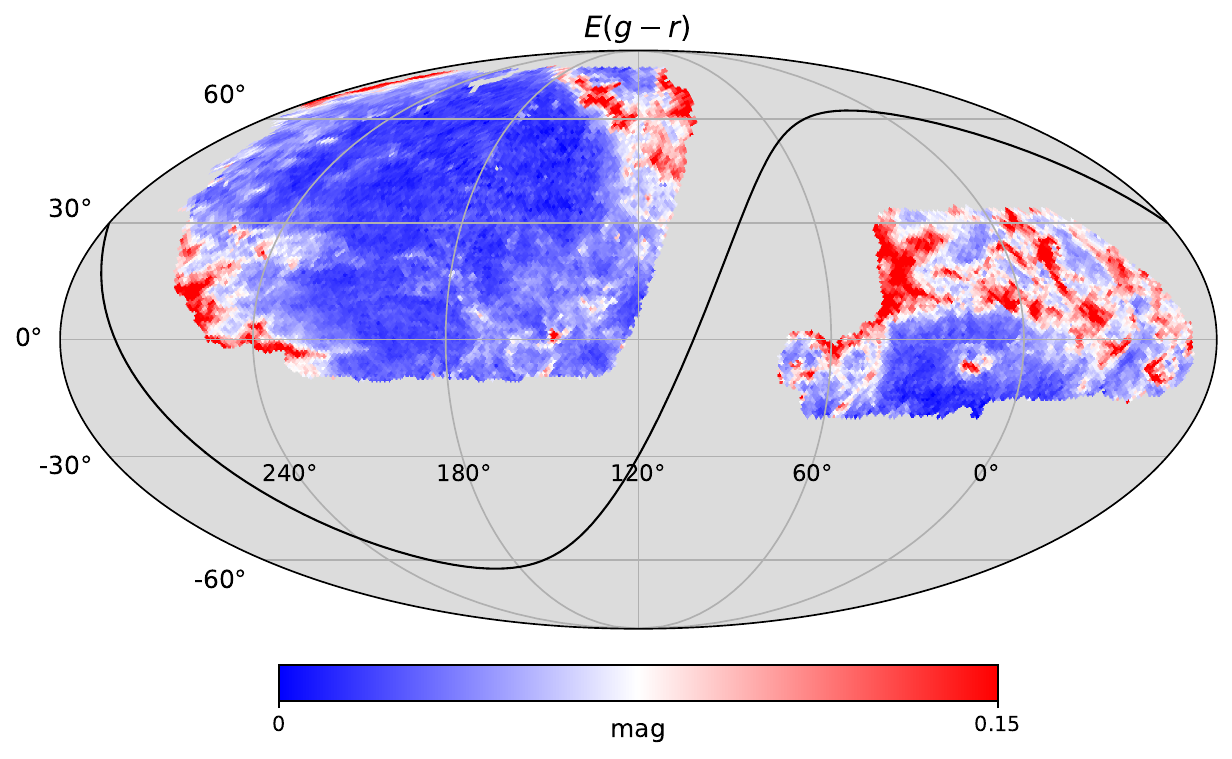}{0.78\textwidth}{(a)}}
    
    \gridline{\fig{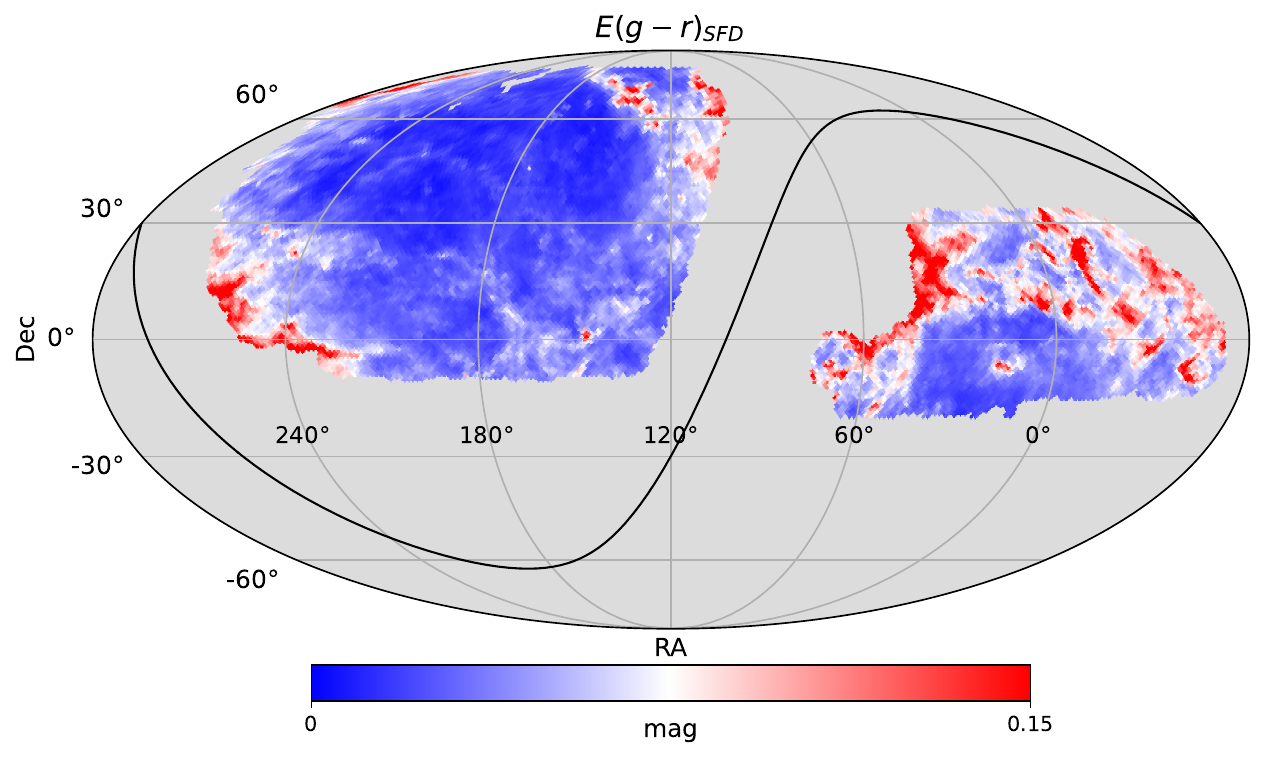}{0.78\textwidth}{(b)}}
    
    \caption{Comparison of $E(g-r)$ maps in celestial coordinates computed via two different approaches---using our measurements derived from DESI data in (a) and from far-infrared thermal dust mapping by SFD in (b). The two maps match closely, though, in detail, significant differences exist. We explore these further in Figure~\ref{fig:rongpu_maps_comparison}(a) and \textsection\ref{subsec:comparison_with_sfd_rongpu}. 
    \textbf{(a)} Reddening in the $g-r$ color, computed by subtracting the intrinsic color predicted by our model from the observed color obtained from the DESI Legacy Imaging Survey. 
    \textbf{(b)} $\ebv$ converted to the reddening in $g-r$ using Equation~\eqref{eq:ebvtoexy}.}
    \label{fig:egr_comparison}
\end{figure*}

A basic test of our new extinction map is to compare it against the widely used SFD map. In Figure~\ref{fig:egr_comparison}b, we compare $\ebv$ to our measured $E(g-r)$, after converting $\ebv$ to the $g-r$ color using Equation~\eqref{eq:ebvtoexy}. Unsurprisingly, the two maps match closely, with a scatter (NMAD) of 8~mmag, compared to typical zero point errors in the DESI Legacy Imaging Surveys of roughly 5~mmag \citep{Dey:2019, Zhou:2023a}. This confirms that our results are in line with the established dust maps in literature.

\begin{figure*}[t]
    \centering
    \gridline{\fig{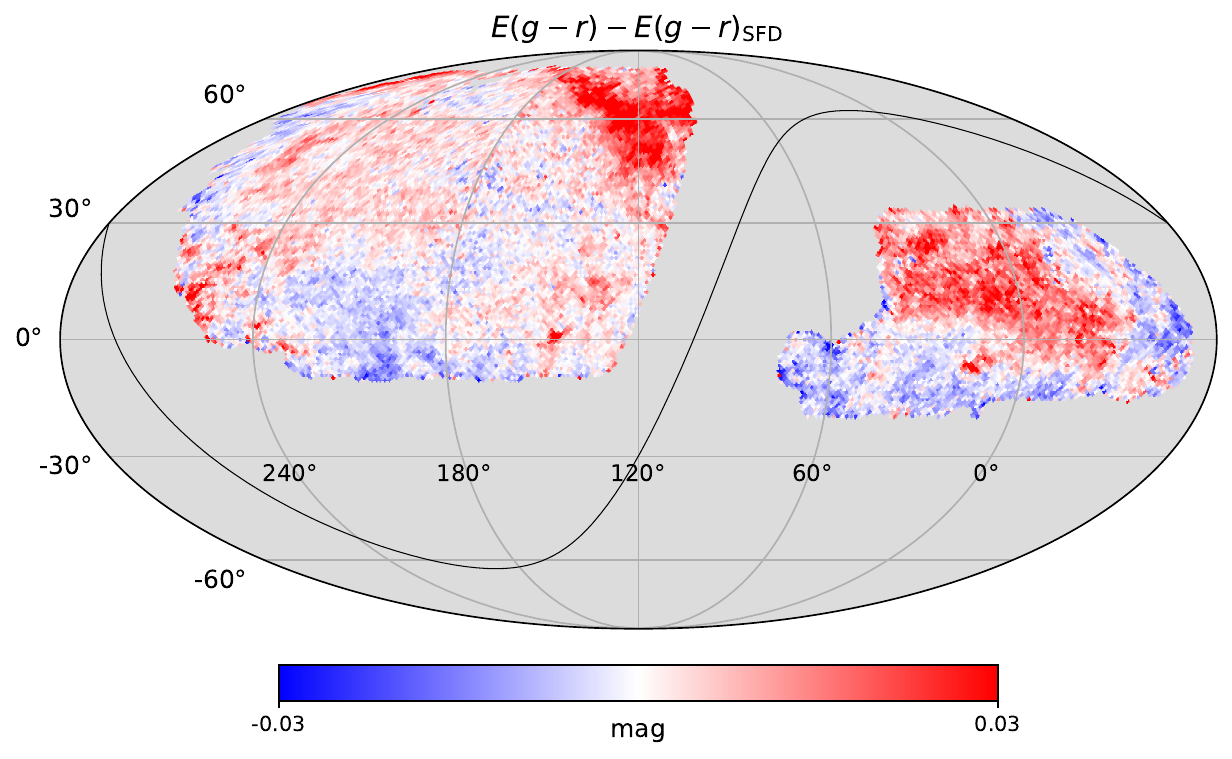}{0.78\textwidth}{(a)}}
    
    \gridline{\fig{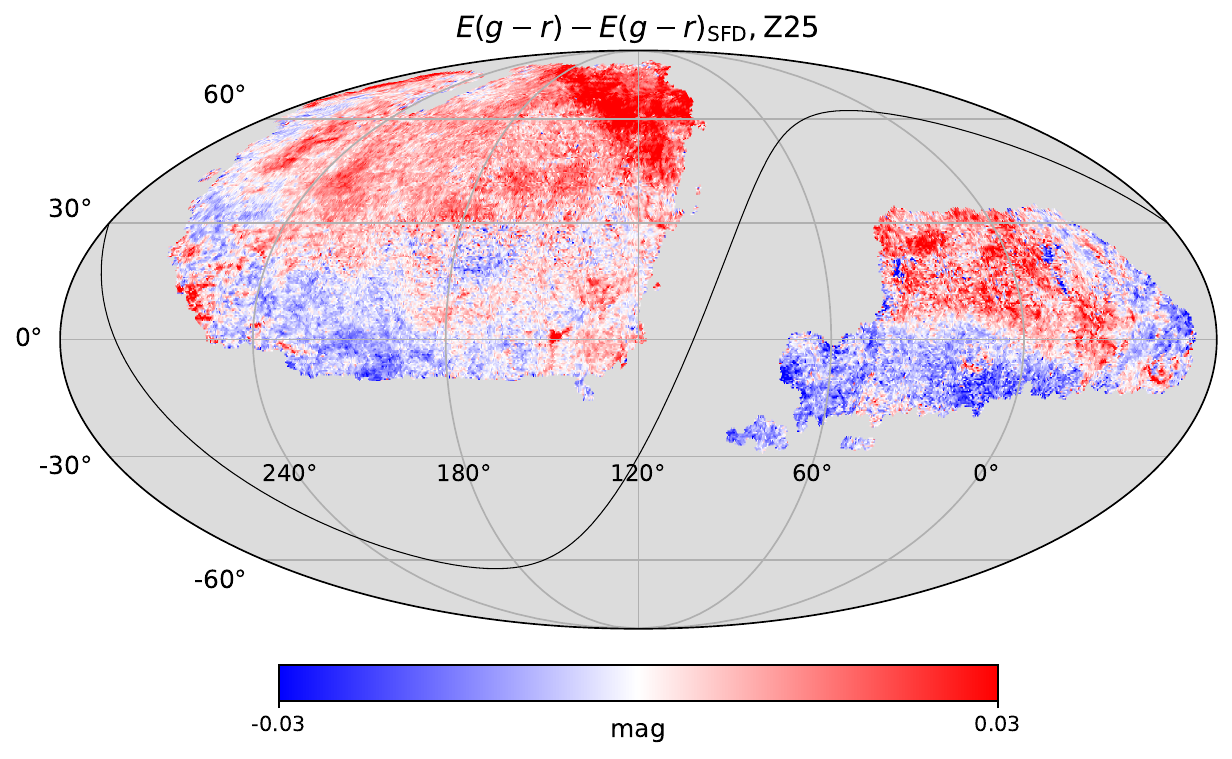}{0.78\textwidth}{(b)}}
    
    \caption{Comparison $E(g-r)$ residuals after subtracting predictions from SFD, using our approach (a) and that of Z25 (b). Note that the two maps cover slightly different footprints by virtue of the different quality cuts and DESI samples used in the two papers. Over the common footprint, the difference between the two maps has a scatter (NMAD) of 6~mmag.  Both maps are in celestial coordinates; our map has a resolution of $55^\prime$, while Z25 has a resolution of $28^\prime$.}
    \label{fig:rongpu_maps_comparison}
\end{figure*}

 While the two maps are broadly consistent, significant and structured residuals are immediately apparent. In Figure~\ref{fig:rongpu_maps_comparison}a, we plot the spatial distribution of the residuals between our $E(g-r)$ and the $E(g-r)$ predicted by SFD. 

We observe large, spatially correlated regions where our measurements are systematically higher (red) or lower (blue) expected from SFD by as much as 0.03 mag. The existence of such large-scale systematic differences indicates that a simple rescaling of the SFD map is insufficient to account for the true dust extinction in these areas. Variations in dust properties impose limitations on the accuracy to which far-infrared reddening corrections can be made, substantially in excess of typical photometric uncertainties.  Direct reddening measurements like ours yield a higher-fidelity extinction map for the DESI survey area.

Next, we compare our work with the DESI reddening map of Z25, which used a similar approach (see \textsection\ref{sec:intro} for differences with our approach). Figure~\ref{fig:rongpu_maps_comparison} shows maps of the difference between $E(g-r)$ computed from DESI data and $E(g-r)_\mathrm{SFD}$ (obtained using Equation~\ref{eq:ebvtoexy}), for both Z25 and this work. The two maps agree at the 6~mmag level (NMAD); unsurprisingly, our related measurements confirm the large spatial scale errors in SFD found by Z25.

\subsection{Consistency in the Reddening Maps in Different Colors}
\label{subsec:egr_erz_egw1}

\begin{figure*}[t]
    \centering
    \gridline{\fig{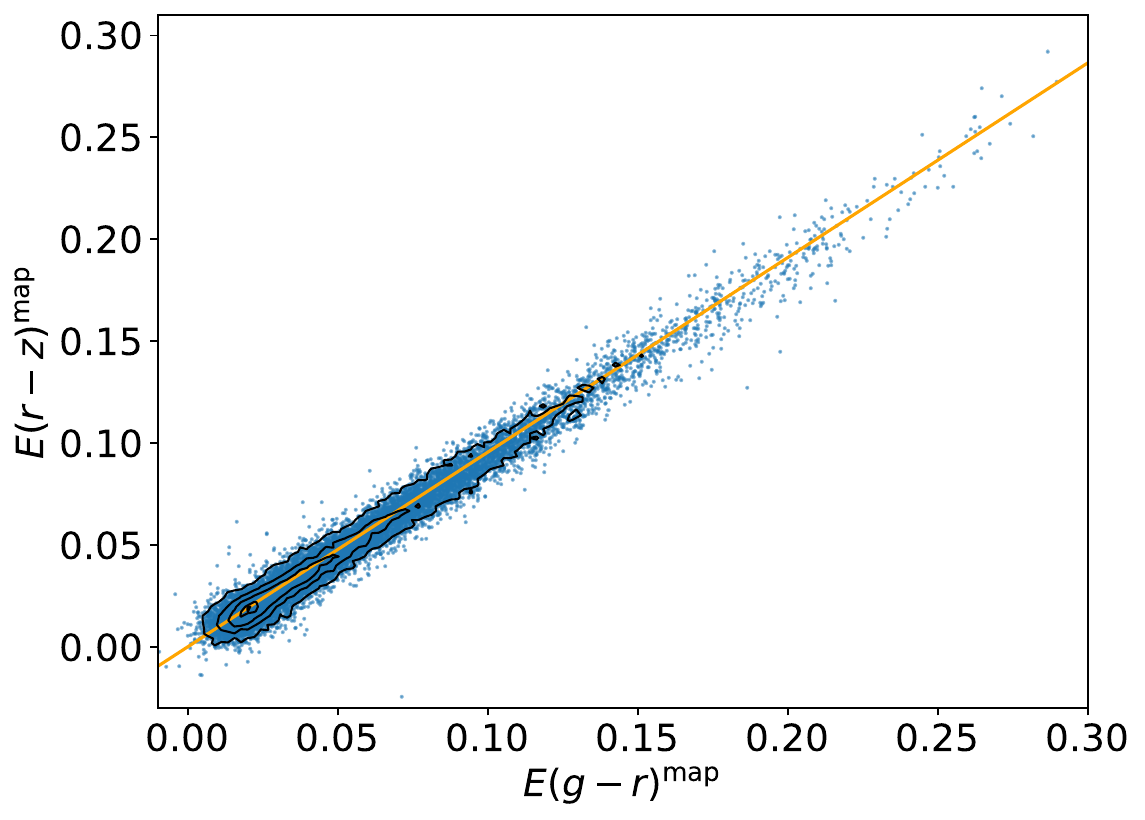}{0.48\textwidth}{(a)}
              \fig{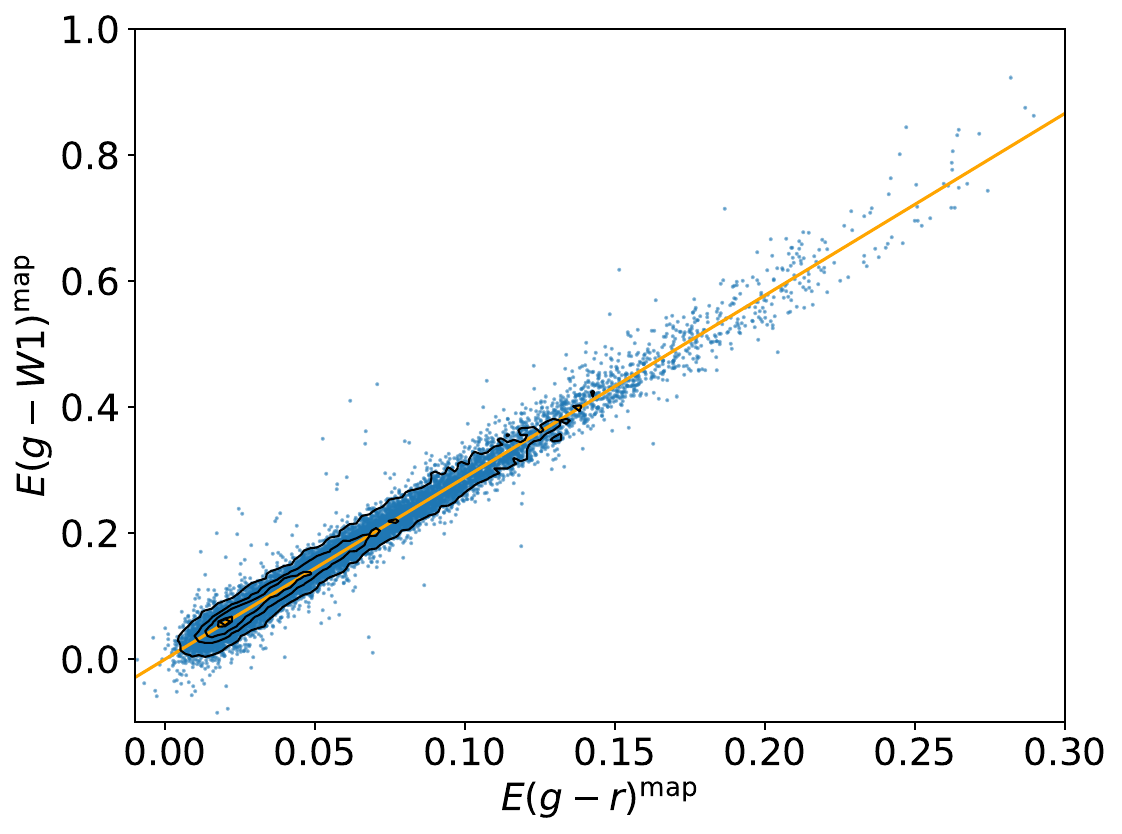}{0.48\textwidth}{(b)}}
    \caption{
        Reddening color-color diagrams derived from map-level residuals. Each point corresponds to a single pixel in our final map. In both panels, the orange line indicates the expected trend for a standard extinction curve.
        \textbf{(a)} The relation between \(E(g-r)\) and \(E(r-z)\). We observe a tight correlation, consistent with expectations from \citealt{Zhou_2025} (see their Figure 14).
        \textbf{(b)} The relation between \(E(g-r)\) and \(E(g-W1)\), which forms the basis of our \(R(V)\) estimate.
    }
    \label{fig:reddening_plots}
\end{figure*}

To test the consistency of our results, we examine reddenings in three colors: $g-r$, $r-z$, and $g-W1$.  The former two colors rely only on DESI Legacy Survey imaging, while the $g-W1$ color gives a longer lever arm, ideal for measuring variation in dust properties via $R(V)$. We show the reddening in $r-z$ and $g-W1$ as a function of the reddening in $g-r$ in Figure~\ref{fig:reddening_plots}.

Unsurprisingly, the reddenings in different colors are tightly correlated.  The slope of the relation is a close match to expectations from the literature \citep{Schlafly:2011} for an $R(V)$ of 3.1.

However,  there is some scatter around the mean relation.  Suggestively, the scatter appears to be largest at the highest reddenings, as would be expected in the presence of variations in the shape of the extinction curve.  We investigate this further in the next section (\textsection\ref{subsec:rv_est_highdust}).

\section{R(V) and Its Variation}
\label{sec:rv_and_variation}

\subsection{Computing $R(V)$}
\label{subsec:rvcomputation}
We now have measurements of reddening in the $g-r$ and $g-W1$ colors.  We would like to relate the reddenings measured in Equation~\eqref{eq:reddening} to $R(V)$, as that is a common measurement of the shape of the extinction curve \citep{CCM:1989}.  In order to do that, as in \cite{Schlafly:2016}, we estimate $R(V)$ as:
\begin{equation}
    R(V) = f\left(\frac{E(g-W1)}{E(g-r)}\right)
\label{eq:rv_est}
\end{equation}
The motivation for this equation comes from Equation~\eqref{eq:rv}.  The $g-r$ color covers a similar part of the optical spectrum as $B-V$, and $A(V)$ is similar to $E(g-W1) = A(g) - A(W1)$ because $A(W1) \ll A(g)$ and $A(g)$ and $A(V)$ are both measurements of the extinction in the optical.  Accordingly, we expect that the relation between $E(g-W1)/E(g-r)$ and $R(V)$ should be relatively simple.

\begin{figure}
    \centering
    \includegraphics[width=\columnwidth]{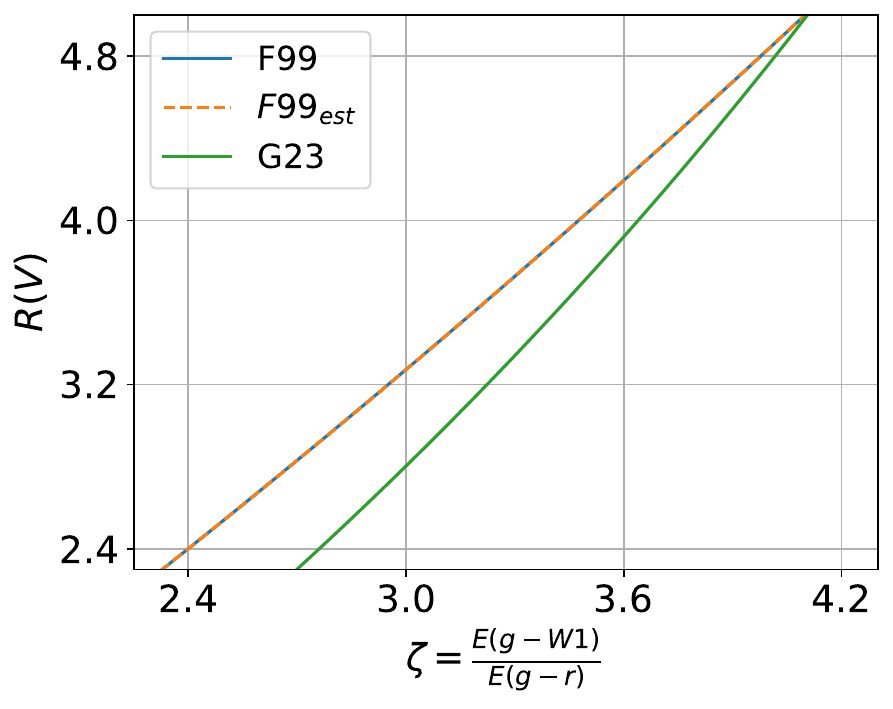}
    \caption{$R(V)$ as a function of the ratio of reddenings in $g-W1$ and $g-r$ colors, for the $F99$ and $G23$ extinction curves. The quadratic fit to $F99$ is given by $\mathrm{F99}_\mathrm{est}=0.07~\zeta^2+1.08~\zeta-0.59$ and is represented by the orange dotted line, which overlays the blue line representing $F99$.}
    \label{fig:rv_plot}
\end{figure}

To compute $f$ in Equation~\ref{eq:rv}, we need to predict extinctions in $g$, $r$, and $W1$ from an extinction curve as a function of $R(V)$, by integrating a stellar spectrum over the relevant bands, as detailed in the appendix of \citet{Schlafly:2011}.  For the spectrum we use a synthetic MARCS spectrum \citep{MARCS:2008} for a model star\footnote{\nolinkurl{p5500_g+4.5_m0.0_t01_st_z-0.75_a+0.30_c+0.00_n+0.00_o+0.30_r+0.00_s+0.00.flx.gz}} with $\teff = 5500 \mathrm{K}$, $\logg = +4.5$, $\feh = -0.75$, close to the typical values of stars studied in this work.  We use a reddening of $A(V) = 0.13$, the median $A(V)$ across the DESI footprint we study.  We use full system throughputs from the DECam and \wise systems.  We compare predictions from both the F99 \citep{Fitzpatrick:1999} and G23 \citep{Gordon:2023} extinction curves.  The resulting reddening ratio is shown as a function of $R(V)$ in Figure~\ref{fig:rv_plot}. As expected, for both the F99 and G23 extinction curves we find that the ratio $E(g-W1)/E(g-r)$ is closely related to $R(V)$, with a slope of roughly 1. However, the two extinction curves provide different estimates of $R(V)$ for a given reddening ratio $\zeta$. For example, for $\zeta = 3.0$, the $R(V)$ values for the two curves differ by about 0.5. We chose to use the F99 extinction curve as opposed to the more recent G23 extinction curve in order to better match past work like \cite{Schlafly:2016}, but we note that the significant disagreement between F99 and G23 at constant $\zeta$ is a major source of systematic uncertainty in converting our measurements to $R(V)$.  

We approximate the numerical relationship between $R(V)$ and $E(g-W1)/E(g-r)$ using a quadratic polynomial to determine $f$, as shown in Figure~\ref{fig:rv_plot}. We then convert our measured reddenings to $R(V)$ using $f$ over our full footprint.

\subsection{$R(V)$ for High-dust Regions}
\label{subsec:rv_est_highdust}

We explore possible variations in $R(V)$ by converting our measured $E(g-r)$ and $E(g-W1)$ maps to $R(V)$ using Equation~\eqref{eq:rv_est}.  The measured $R(V)$ are noisy at low reddening because they involve the ratio of two reddenings, so we focus on the $E(g-r) > 0.1$ regime where the reddenings are approximately 20 times larger than the expected photometric uncertainties in the $E(g-r)$ color \citep{Dey:2019, Zhou:2023a}.  

\begin{figure*}[t]
    \centering
    \gridline{\fig{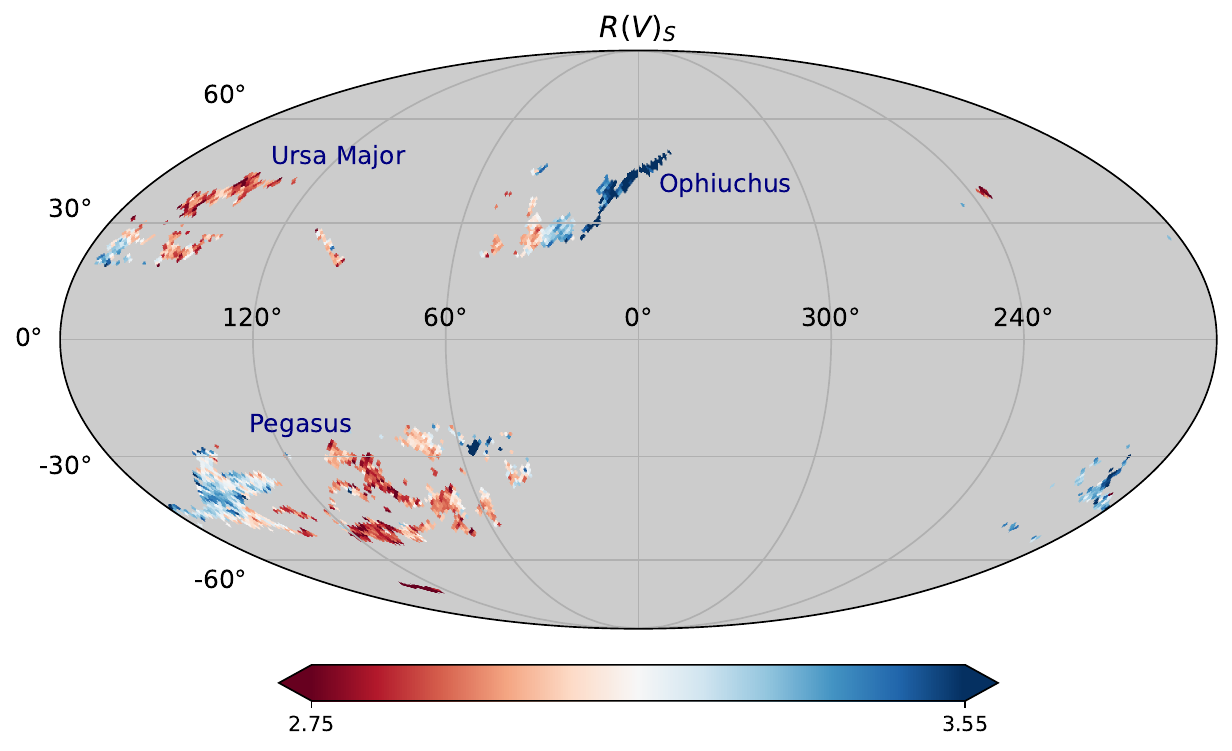}{0.78\textwidth}{(a)}}
    
    \gridline{\fig{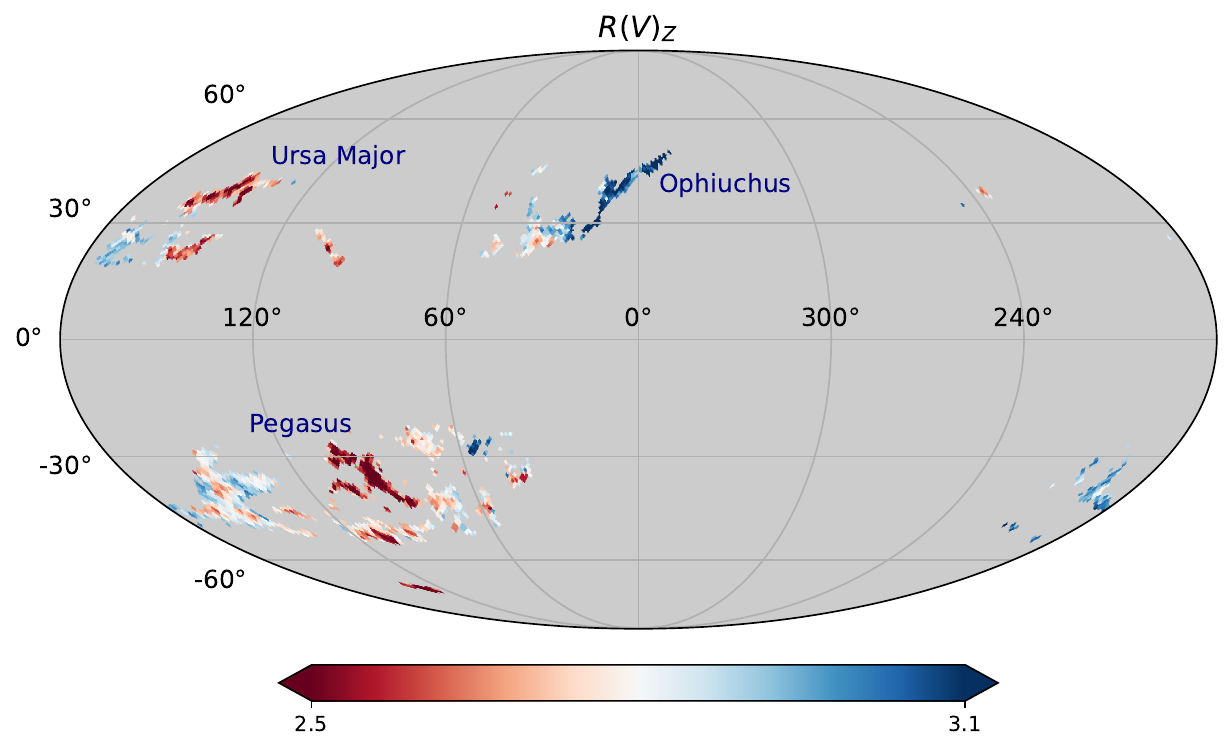}{0.78\textwidth}{(b)}}
    
    \caption{Comparison of two different $R(V)$ maps in Galactic coordinates over the DESI footprint for high-dust regions. Note that the $R(V)$ color scales are different, but the variations within the two maps are strongly correlated. 
    \textbf{(a)} $R(V)$ results from this paper, over the DESI footprint with $E(g-r) > 0.1$. $R(V)$ has a median value of 3.1, with significant correlated variations across the sky. 
    \textbf{(b)} $R(V)$ from \cite{Zhang&Green:2024} using cells with distances beyond 1.5~kpc. The close correlation with our $R(V)$ map is evident, tracing similar highs in Ophiuchus and lows in Pegasus and Ursa Major.}
    \label{fig:rv_maps_comparison}
\end{figure*}

Figure~\ref{fig:rv_maps_comparison}(a) shows the spatial variation of our estimated $R(V)$ in regions with $E(g-r) > 0.1$.  We see significant variations in $R(V)$ in this footprint, with highs of $R(V) = 3.4$ in the vicinity of Ophiuchus and lows of $R(V) = 2.9$ over portions of the south and north Galactic caps.  

These measured $R(V)$ variations seem robust: we do not see strong trends with $E(g-r)$ in these regions (see \textsection\ref{subsec:ebv_rv}), as might be expected for errors in the photometric calibration, and the maps show correlations over large spatial scales.

We can further confirm these results by comparing with the \gaia XP-based $R(V)$ measurements of \cite{Zhang&Green:2024}\footnote{We use the associated data product at \url{https://zenodo.org/records/11394477}}, hereafter ZG24.  We compute a high-latitude $R(V)$ map from their 3D $A(B)$ and $A(V)$ maps by averaging together all slices of their 3D maps with heliocentric distances greater than 1.5 kpc\footnote{Averaging these maps essentially uses more \gaia stars to increase the signal to noise of the $B$ and $V$ extinction measurements.  Note that these maps are cumulative \emph{A(B)} and \emph{A(V)} maps, measuring the \emph{total} extinctions to the stars at these distances; they include all dust more nearby than 1.5~kpc.}.  Specifically, we average together their measurements of the total extinctions in the B and V bands in each pixel in these slices, and then construct $R(V) = A(V) / (A(B) - A(V))$.  The distance of 1.5 kpc corresponds to 500 pc above the plane at $\lvert b \rvert = 20\degree$, significantly larger than the dust scale height of 100 pc, meaning that the derived average $R(V)$ map includes almost all of the dust that will be present in the DESI footprint.

As shown in Figure~\ref{fig:rv_maps_comparison}, we find good agreement with the ZG24 results---the principal structures in the spatial distribution of $R(V)$ are remarkably similar between the two methods.  Both find the highest $R(V)$ (e.g., Ophiuchus, north of the Galactic center) and lowest $R(V)$ (e.g. Ursa Major and Pegasus) in the same areas. 

This agreement is particularly compelling given the very different technique used to estimate $R(V)$ in ZG24. They simultaneously model the stellar type, extinction, and extinction curve shape using the \gaia XP low-resolution ($R < 100$) spectrophotometry, without having access to relatively high-resolution DESI spectroscopy ($R \approx 4000$), which we use to determine the spectral types of stars independently of their reddenings.

\begin{figure}
    \centering
    \includegraphics[width=\columnwidth]{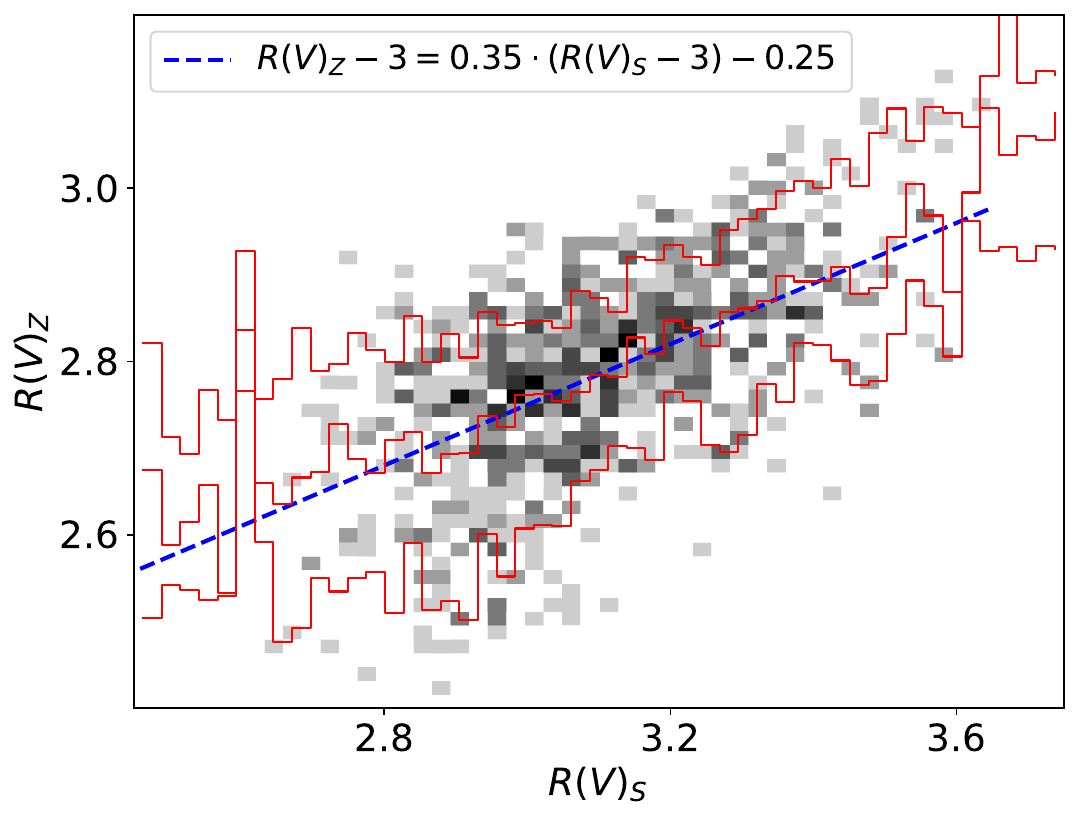}
    \caption{A comparison of our $R(V)$ with $R(V)$ computed by ZG24. The two maps are strongly correlated, though they differ significantly in both scale and offset. Note that these maps are in celestial coordinates unlike the maps in Figure~\ref{fig:rv_maps_comparison}.}
    \label{fig:rv_zg vs rv_ss}
\end{figure}

That said, while the $R(V)$ maps of Figure~\ref{fig:rv_maps_comparison} are strikingly similar, the two maps have a significantly different scale and offset.  Our map has a median $R(V)$ of 3.1 with a normalized median absolute deviation of 0.23, compared to 2.8 and 0.16 respectively for the map of ZG24. In Figure~\ref{fig:rv_zg vs rv_ss} as a scatterplot, transforming the ZG24 map to celestial coordinates, confirming the good correlation between the maps.  We note that for $\rvs$ (our measurements) of 3.0, $\rvz$ (the measurements of ZG24) is 2.75. Further, the variation in $R(V)_Z$ is about a third (0.35) of the variation in $R(V)_S$ that we measure. The overall offset in $R(V)$ is perhaps not surprising---for example, Figure~\ref{fig:rv_plot} shows that significant differences in the overall $R(V)$ occur as a result of choosing different extinction curves from the literature.  The different scale of the variation in $R(V)$ is harder to understand; we posit that this may be due to the greater importance of priors on $R(V)$ in the work of ZG24 due to their need to simultaneously model both the stellar type and extinction curve of the individual stars.  Moreover, because a significant fraction of the points on our map lie near masked, $E(g-r) < 0.1$ regions, the choice of coordinate system we use to make the comparison makes a modest difference to the slope (0.35 -- 0.55).  Still, the strong correlation between the two independent measurements of $R(V)$ shown in Figure~\ref{fig:rv_zg vs rv_ss} serves to demonstrate the reliability of both sets of measurements.

\subsection{Variation of $R(V)$ against E(B-V)}
\label{subsec:ebv_rv}
\begin{figure}
    \centering
    \includegraphics[width=\columnwidth]{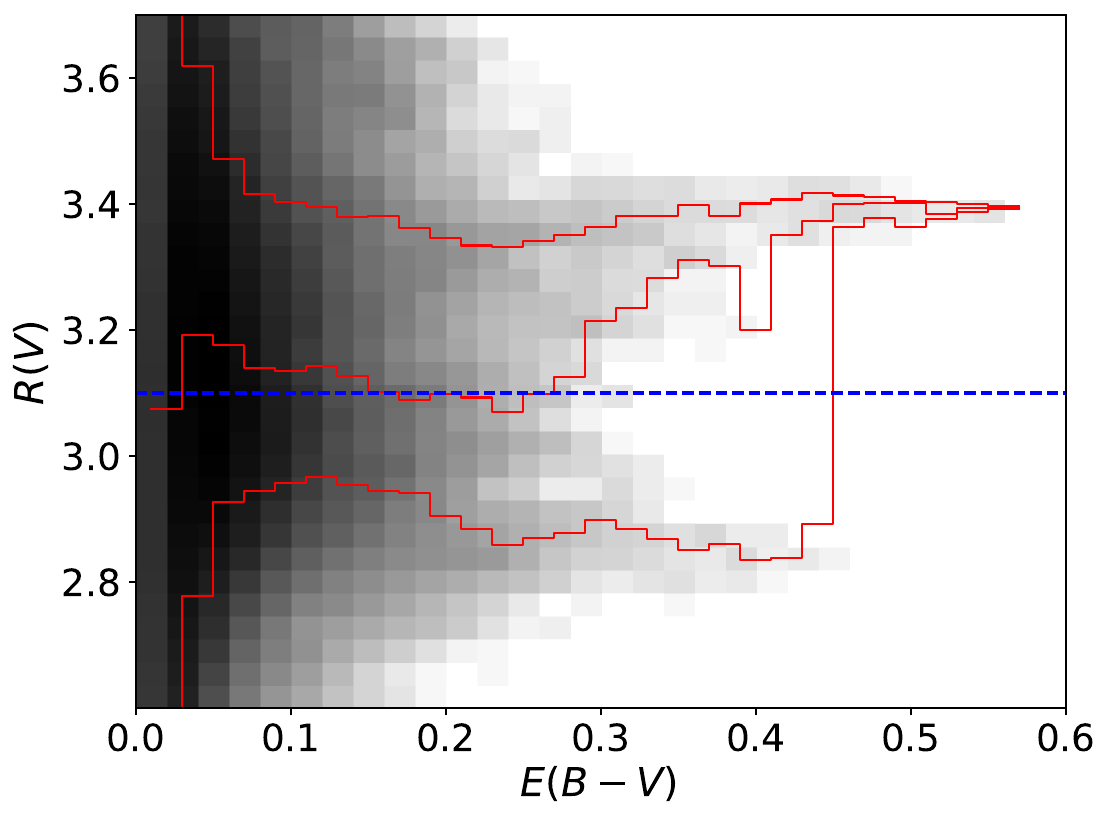}
    \caption{$R(V)$ as a function of $E(B-V)$. At low reddenings, the $R(V)$ values, computed as a ratio of reddenings are very uncertain. For $E(B-V) > 0.2$, one can see a clear bimodality; i.e., there are two dark gray peaks in $R(V)$, centered around 2.9 and 3.4. The blue dashed line represents $R(V)=3.1$, the average value across diffuse interstellar medium.}
    \label{fig:rv_vs_ebv}
\end{figure}

Many processes in grain evolution depend on the density of dust grains \citep{Draine:2003}. In Figure~\ref{fig:rv_vs_ebv}, we investigate whether $R(V)$ depends on dust column as traced by SFD, a rough proxy for dust density.  We find that in the high signal-to-noise regime ($E(B-V) > 0.2$), the distribution of $R(V)$ becomes notably structured. The distribution of $R(V)$ is clearly bimodal, with the data clustering around two preferred values: $R(V) \approx 2.9$ and $R(V) \approx 3.4$. It is unclear what causes this bimodality, but we conservatively posit that it is due to the relatively small number of significant dust clouds that overlap the DESI footprint and the large scale correlations in $R(V)$ across these clouds---given the large scale correlation in $R(V)$, we simply do not have many independent draws from the range of possible dust $R(V)$ in the DESI footprint.  At low reddening ($E(B-V) \lesssim 0.1$), the $R(V)$ measurements are dominated by noise.

\section{Discussion}
\label{sec:discussion}

\subsection{Potential Causes of $R(V)$ variation}
\label{subsec:rv_variation_causes}

Variations in $R(V)$ are often linked to the size distribution of dust grains, with larger average grain sizes producing higher $R(V)$ \citep{Draine:2003}. Other factors that influence the extinction ratio include dust composition, grain shape, and environment \citep{Jones:2013, Mulas:2013}. To try to understand the potential causes of its variation, we look at a variety of probes of the interstellar medium and its environment.

\begin{figure*}
    \centering
    \includegraphics[width=0.93\textwidth]{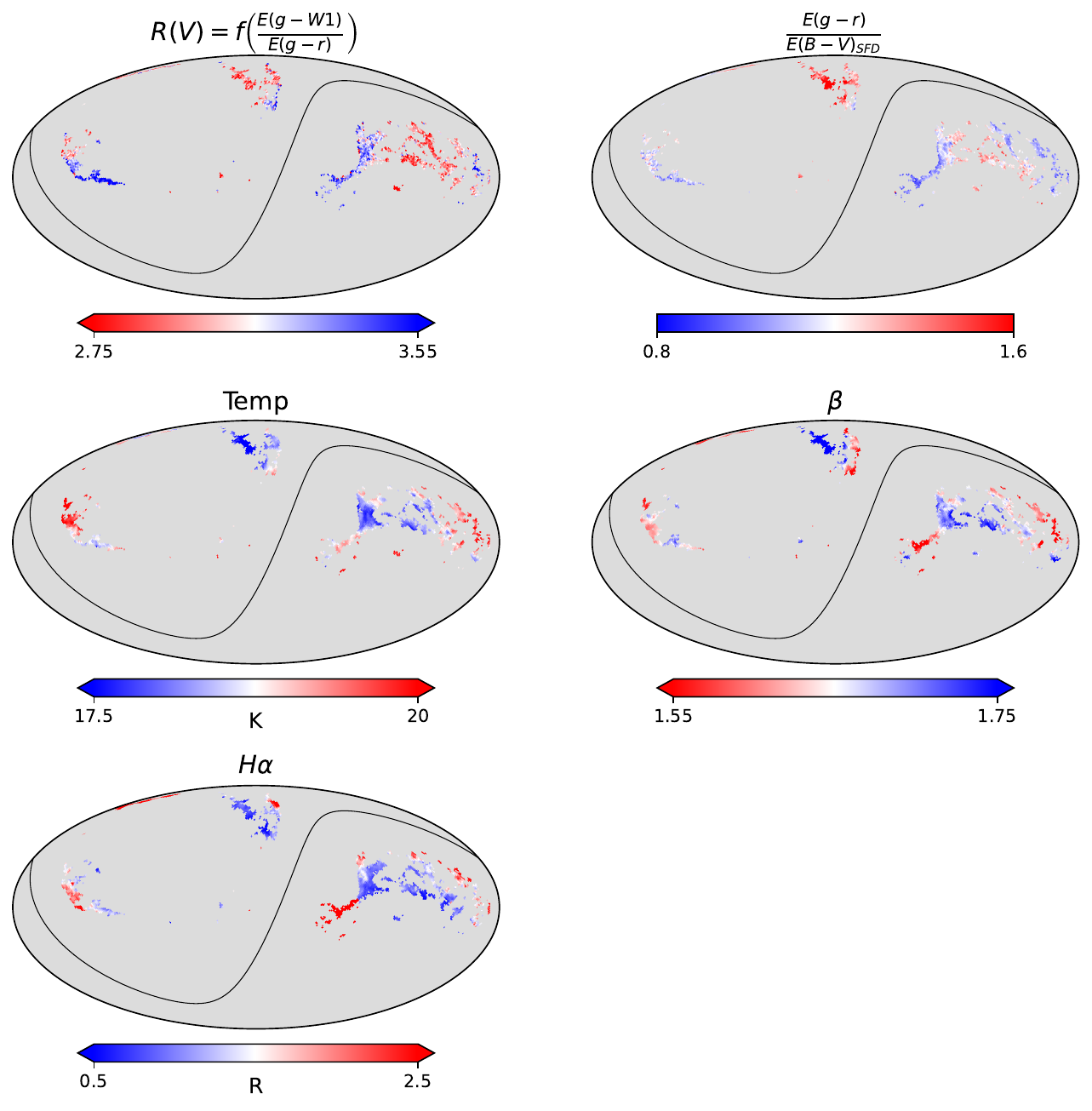}
    \caption{Comparison of different tracers for extinction curve variations and physical dust properties across the sky. Some panels show clear spatial correlations, indicating they are tracing related physical phenomena. All maps are in celestial coordinates. \newline\textbf{Top Left:} $R(V)$ from photometric reddening residuals $E(g-W1)$ and $E(g-r)$. This map highlights deviations from the average Galactic extinction curve. \newline\textbf{Top Right:} The reddening residual in $g-r$ compared to the $E(B-V)$ obtained from SFD. This shows regions where we find different amounts of extinction than what is predicted from the SFD map. Its strong anti-correlation with the top left panel indicates that extinction curve variations drive errors in the SFD prediction. Note that this color scale is flipped as compared to that of $R(V)$. \newline\textbf{Mid Left:} The dust temperature, as measured by the \planck satellite. \newline\textbf{Mid Right:} The dust emissivity power-law index ($\beta$), also from \planck. The spatial patterns in $\beta$ are visually similar to the reddening residuals in the top two panels, while the temperature and $\beta$ maps are anti-correlated. Note that the color scale of $\beta$ is flipped in comparison to that of dust temperature.
    \newline\textbf{Bottom Left:} $H\alpha$ map depicting ionizing radiation. It correlates well with $\beta$ and $T$, and poorly with $R(V)$.}
    \label{fig:rv_correlation_maps}
\end{figure*}

Figure~\ref{fig:rv_correlation_maps} shows several maps related to dust and the interstellar medium.  The top left panel shows our $R(V)$ map for comparison with the other panels. In the top right panel, we show the ratio between the observed $g-r$ reddening and the SFD-predicted reddening in $B-V$. It shows a strong anti-correlation with $R(V)$. This corroborates our measurements: the areas where SFD predicts the wrong amount of extinction are the same areas where we find that the extinction curve differs from the average extinction curve.  It is also in some ways unsurprising: SFD uses a single average prescription to translate the thermal brightness of dust into extinction, but we find that the extinction curve varies, so the single average prescription cannot possibly make good predictions of the extinction over the full footprint. 

\begin{figure}[t]
    \centering
    \gridline{\fig{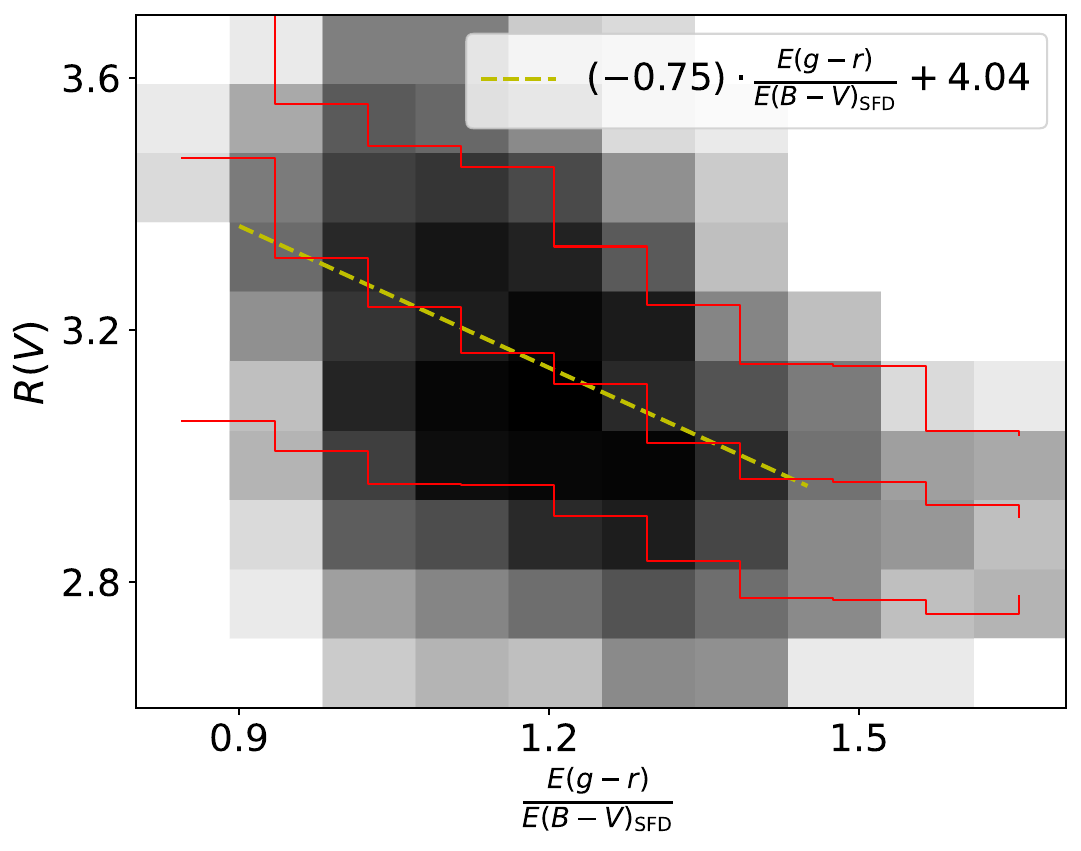}{0.9\linewidth}{(a)}}
    
    \gridline{\fig{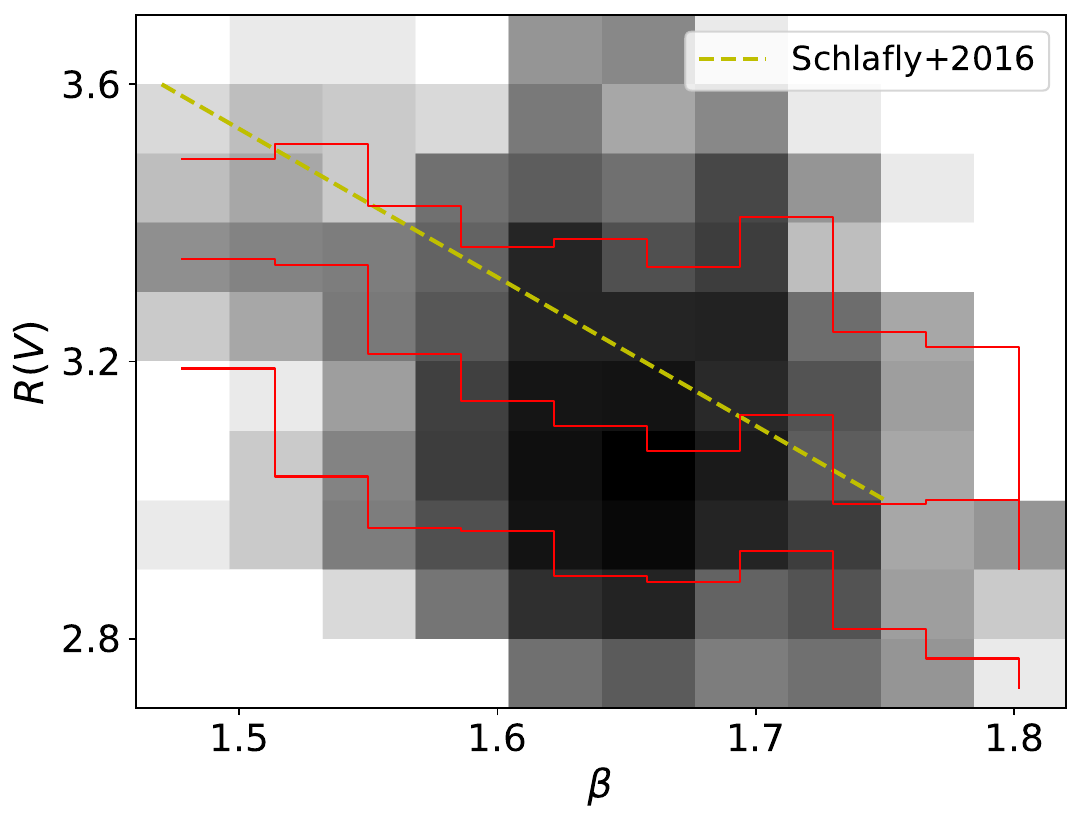}{0.9\linewidth}{(b)}}
    
    \caption{
        Comparison of $R(V)$ with $E(g-r)/E(B-V)_\mathrm{SFD}$ (a) and $\beta$ (b).  The three red lines depict the 16th, 50th, and 84th percentiles.  There is a good correlation between $R(V)$ and $E(g-r)/E(B-V)_\mathrm{SFD}$, perhaps unsurprisingly because SFD uses an average extinction curve to convert from FIR measurements to optical extinctions.  There is only a mild correlation between $R(V)$ and $\beta$ at high Galactic latitudes, not as striking as that found by \cite{Schlafly:2016} for low Galactic latitudes.  The \cite{Schlafly:2016} relation is shown as the dashed line.}
    \label{fig:rv_scatter_plots}
\end{figure}

We show the relationship between our measured $R(V)$ and $E(g-r)/E(B-V)_\mathrm{SFD}$ in Figure~\ref{fig:rv_scatter_plots}(a), which are well anti-correlated.  An important corollary of this correlation is that the combination of far-infrared-based measurements of the dust extinction and a single measured optical reddening provides a useful handle on $R(V)$.  Armed only with measurements of $E(g-r)$ and SFD, Figure~\ref{fig:rv_scatter_plots}(a) indicates that it is possible to estimate $R(V)$ with a precision of $\sim 0.14$, at least at high Galactic latitudes. This result allows taking advantage of the high signal-to-noise ratios available in the far-infrared dust emission measurements to measure $R(V)$, even for surveys without the large optical-infrared lever arms needed to estimate $R(V)$ directly from the extinction curve.

Grain growth through coagulation and mantle formation is expected to be most efficient in cold, dense environments, affecting the grain size distribution and $R(V)$ \citep{Whittet:1988, Ysard:2013}. While this effect is expected at higher densities than we probe here, it motivates studying whether the $R(V)$ variations we observed are correlated with dust density or dust temperature.  We already observed no correlation between column density and $R(V)$ in Figure~\ref{fig:rv_vs_ebv}. The mid left panel in Figure~\ref{fig:rv_correlation_maps} shows the dust temperature map from \cite{Planck:2016}\footnote{\url{https://irsa.ipac.caltech.edu/data/Planck/release_2/all-sky-maps/maps/component-maps/foregrounds/COM_CompMap_Dust-GNILC-Model-Temperature_2048_R2.00.fits}} for comparison with our map of $R(V)$. While there are some areas where the two maps have similar trends, there is no strong correlation over the entire footprint.

Continuing to investigate possible correlations with $R(V)$, we now consider the far-infrared dust emissivity spectral index $\beta$, as measured by \cite{Planck:2016}\footnote{\url{https://irsa.ipac.caltech.edu/data/Planck/release_2/all-sky-maps/maps/component-maps/foregrounds/COM_CompMap_Dust-GNILC-Model-Spectral-Index_2048_R2.00.fits}}.  The work of \citet{Schlafly:2016} at low Galactic latitudes previously reported a correlation between $\beta$ and $R(V)$, suggesting that the far-infrared spectral energy distribution of dust may predict $R(V)$.  Both $R(V)$ and $\beta$ may depend on the details of the composition and chemistry of dust grains, making this correlation plausible, though existing models do not naturally produce a correlation \citep{Zelko:2020}.  Figure~\ref{fig:rv_scatter_plots}(b) shows a plot of $R(V)$ against $\beta$ for our data (at high Galactic latitudes) along with a line representing the data for a similar plot from \cite{Schlafly:2016}. We do not see great agreement, although there is a similar trend and a mild correlation. In the mid right panel of Figure~\ref{fig:rv_correlation_maps}, we show a map of $\beta$.  At high Galactic latitudes, dust temperature and $\beta$ are strongly anticorrelated.  It is unclear if this is due to systematic uncertainties in the $\beta$ measurements at high latitudes with relatively low signal levels or due to other differences between nearby, diffuse dust clouds and more dense material at lower latitudes.

Energetic radiation shapes dust grains through a variety of processes, including photo-evaporation of dust mantles, destruction of polycyclic aromatic hydrocarbons, and UV photo-processing of grain surfaces \citep{Draine:2003, Jones:2013}.  We compare our map of $R(V)$ with a map of $H\alpha$, as produced by \citet{Finkbeiner:2003}\footnote{\url{https://lambda.gsfc.nasa.gov/data/foregrounds/fink_halpha/Halpha_fwhm06_1024.fits}},  shown in the bottom left panel of Figure~\ref{fig:rv_correlation_maps}, which probes the amount of ionized gas along each line of sight.  We find only that $H\alpha$ is well correlated with dust temperature and anticorrelated with $\beta$.  None of these three quantities correlate well with our $R(V)$ maps suggesting that these are not driving the observed variation in $R(V)$.

\subsection{Implications for Far-Infrared-based Dereddening}
\label{subsec:farIR_dereddening}

Extragalactic photometry is typically dereddened using a far-infrared dust
map together with a single extinction curve. The relation in
Figure~\ref{fig:rv_scatter_plots}(a) lets us estimate the error this
introduces. For a scatter in $R(V)$ of 0.23, the implied scatter in
$E(g-r)$ at fixed $E(B-V)_{\rm SFD}$ is

\begin{align}
\sigma(E(g-r)) &= \frac{\sigma(R(V))}{0.75}\,E(B-V)_{\rm SFD} \nonumber \\
&\approx 0.31\,E(B-V)_{\rm SFD} \, .
\label{eq:fir_egr}
\end{align}
This is a steep relationship and implies a 30\% uncertainty on \ebv\ at low dust columns, putting a high floor on the accuracy to which FIR dereddening can be performed on cosmological footprints.  For example, most DESI cosmology analyses focus on low reddenings.  Focusing on $E(B-V) < 0.1$, the median $E(B-V)_\mathrm{SFD}$ is 0.033~mag.  At that reddening, this relation implies a 1\% photometric accuracy floor due to dereddening uncertainties, which matches the typical scatter we and Z25 see when comparing $E(g-r)$ and $E(B-V)_\mathrm{SFD}$.  This is higher than the typical level of systematic uncertainty associated with SFD (e.g., \citealt{Schlafly:2010}).  This estimate relies on extrapolating the observed dispersion in $R(V)$ at $E(B-V) > 0.1$ to lower columns, which seems like the right assumption but which is hard to verify in practice.

\section{Systematic Uncertainties}
\label{sec:sys}

\subsection{The Uncertainty Floor}
\label{subsec:uncfloor}

We obtain model residuals of 17~mmag in $g-r$ and 77~mmag in $g-W1$ in our calibration sample, as shown in Table~\ref{tab:locally_corrected_residuals}.  We would like to know the source of these errors in order to guide future work measuring $R(V)$ at lower $E(B-V)$ and higher precision.

The work of \citet{Zhou:2023a} finds a systematic uncertainty floor on the DESI Legacy imaging of 4--5~mmag, significantly smaller than our error floor, though isolated regions can show residuals up to $\sim20$~mmag.  We conclude that the photometric calibration is only a small contributor to the per-star residuals.

We see trends in the amount of scatter in the residuals with spectroscopic signal-to-noise and photometric magnitude, and so construct a high signal-to-noise subsample in order to assess the limiting per-star color accuracy we can reach.  We select stars with $SN_B > 50$, $16.5 < g < 17.5$, $\mathrm{W1} < 18$, and follow the procedure of Equation~\ref{eq:small_scale_model_residual} to remove photometric calibration errors or correlated dust extinction errors from the model residuals.  We achieve an accuracy of $\sigma_{\rm{NMAD}}(\delta_{\rm g-r}) = 12$~mmag and $\sigma_{\rm{NMAD}}(\delta_{\rm g-W1}) = 49$~mmag, similar to the best performance obtained in the analogous work of Z25.  This is close to the typical percent-level systematic floor often invoked for bright star photometry in $g-r$ but is rather implausibly far from any photometric limit in $\mathrm{W1}$.  Interestingly, we find that the residuals in $g-\mathrm{W1}$ are decently correlated with the residuals in $r-z$ in this sample.  We can further reduce the scatter in $g-\mathrm{W1}$ by subtracting a scaled version of the $r-z$ residuals, from 49~mmag to 35~mmag, reducing the variance by a factor of two.  We expect that this means that our per-star systematic uncertainty floor is dominated by uncertainties in the derived spectroscopic parameters, and in particular the temperatures, which contribute correlated noise to all colors.

\needspace{4\baselineskip}
\subsection{Errors in Predicted Colors Due to Imperfect Calibration Star Dereddening}
\label{subsec:syslowebvcolors}
Systematic uncertainties in our predicted colors due to residual reddening in our low-reddening calibration sample will limit the accuracy to which we can measure $R(V)$.  We use SFD-deredenned colors of low-reddening ($E(B-V) < 0.04$~mag) stars to train our model of the intrinsic colors of stars and apply this model over the whole footprint; errors in these intrinsic colors will imprint on our $R(V)$ maps.  We assess the importance of this systematic by separately training a model on the same low-reddening calibration stars, but where instead of deredenning by SFD, we have dereddened by a constant $E(B-V) = 0.03$~mag, roughly matching the average value of SFD on the $E(B-V) < 0.04$~mag footprint.  We find that we obtain consistent intrinsic colors in $g-r$ and $g-W1$ at the 1.3 and 3.9~mmag level, after removing a small north/south difference.  These differences are more than ten times smaller than our typical per-star residuals (Table~\ref{tab:locally_corrected_residuals}).

We further constructed maps of $R(V)$ from the alternatively constructed intrinsic color models.  For the sky footprint with $E(B-V) > 0.1$, our usual cut for analyzing $R(V)$, we obtain maps different by only $\sigma_{\mathrm{NMAD}}(\Delta R(V)) = 0.005$, much smaller than our observed variation in $R(V)$ of 0.23.

These results give us confidence that our measurements are minimally affected by our treatment of the dereddening of the calibration stars.  We note also that in this test we are performing essentially \emph{no} dereddening correction of the calibration stars and are primarily relying on the fact that the stars' stellar parameters are uncorrelated enough with the dust reddening at high latitudes that the intrinsic color model just does not have the flexibility to fit out dust-driven changes in observed color.  Accordingly, the results here are a conservative upper bound on this source of systematic uncertainty.

\subsection{Systematic Uncertainty in Intrinsic Color Modeling}
\label{subsec:color_model_systematics}
We choose in \textsection\ref{subsec:colormodel} to model intrinsic colors as a function of the temperature, metallicity, and surface gravity of stars.  In Equation~\eqref{eq:colormodel}, $\teff$ is the most influential parameter, followed by $\feh$ and $\logg$. We explore here whether expanding that model to additionally include $[\mathrm{\alpha/Fe}]$ makes an important difference to our analysis. Training with $[\mathrm{\alpha/Fe}]$ as an additional feature improves the normalized median absolute deviation for $g-r$ and $g-W1$ by $\le$1~mmag. The resulting $R(V)$ maps for the sky footprint with $E(B-V) > 0.1$ using these augmented models differ from our presented results by only $\sigma_{\mathrm{NMAD}}(\Delta R(V)) = 0.015$, much lower than the observed scatter of 0.23 (\textsection\ref{subsec:rv_est_highdust}). These more detailed measurements of the chemistry of stars do not affect the intrinsic colors of stars enough to impact the measurements of $R(V)$ in this work meaningfully.

\subsection{Uncertainty in $R(V)$ measurement on account of stellar spectrum}
\label{subsec:spectrum_effect_on_rv}
\begin{figure}
    \centering
    \includegraphics[width=\columnwidth]{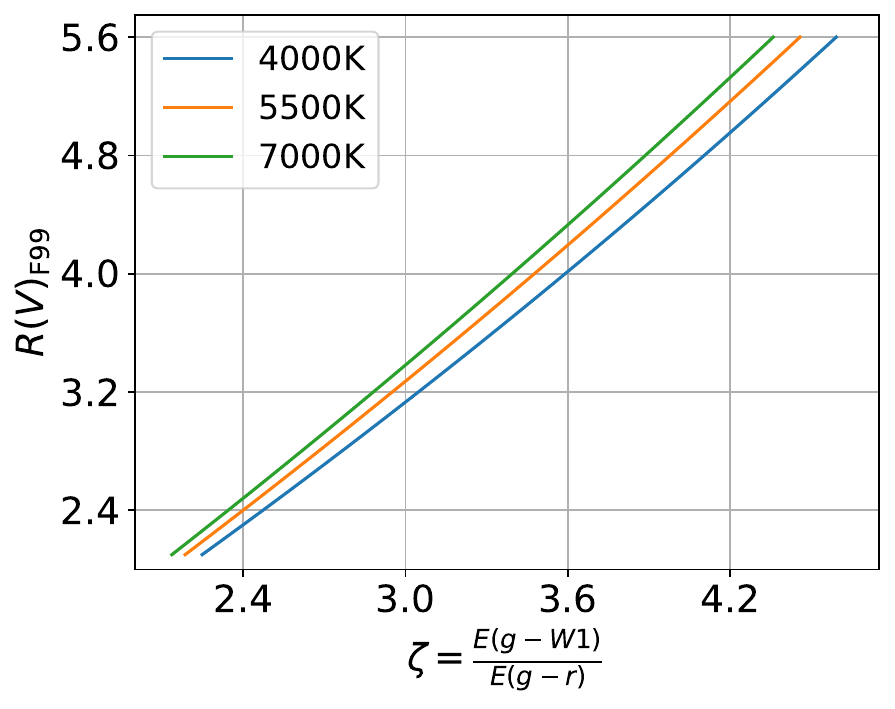}
    \caption{$R(V)$ obtained using \cite{Fitzpatrick:1999} against the ratio of $E(g-W1)$ and $E(g-r)$. Using spectra from synthetic stars with different temperatures but same values for all other spectroscopic parameters results in a different value of $R(V)$ for a given reddening ratio. 4000K and 7000K are the minimum and maximum stellar temperatures in our sample, while 5500K is a typical value.}
    \label{fig:rv_vs_zeta}
\end{figure}

The function $f$ in Equation~\ref{eq:rv_est} depends slightly on the spectrum of the star, which we neglect in our analysis. In Figure~\ref{fig:rv_vs_zeta}, we analyze how this relation depends on the choice of spectrum, using the minimum and maximum temperature stars we consider. At $\zeta = 3$, we see that $R(V)$ varies by $\sim0.1$ for a variation of $\Delta \teff$ of 1000~K. The mean $\teff$ per pixel varies across our footprint with an NMAD of $\sim$100~K, meaning that the scatter in effective temperature of stars results in a typical error of $\sim$0.01 in $R(V)$. The effect of the variation in the average values of other spectroscopic parameters ($\feh$, $\logg$) over the footprint is smaller still. Thus, we neglect the variation in $R(V)$ with stellar type in this work.

\section{Data Availability}
\label{sec:data_availability}
The data product for this work is available at \dataset[doi:10.5281/zenodo.23151436]{https://doi.org/10.5281/zenodo.23151436}. We provide reddening maps (\texttt{nside} = 64, RING ordering,  celestial coordinates) of our measurements in $g-r$, $r-z$, and $g-W1$ colors, along with $E(B-V)_{\rm SFD}$ for reference. We also provide a map of $R(V)$, derived from $E(g-W1)/E(g-r)$ via Equation~\eqref{eq:rv_est}, and a map of $E(g-r)/E(B-V)_{\rm SFD}$. We suggest using the latter two maps only for pixels with $E(g-r) > 0.1$ following the discussion in \textsection\ref{subsec:rv_est_highdust}. We recommend the $E(g-r)$ map for general reddening corrections, as it has the least residual scatter (Table~\ref{tab:locally_corrected_residuals}). A script for reproducing the figures derived from our $R(V)$ maps and related products is also included in the data product.

\section{Conclusion}
\label{sec:conclusion}

We present a method for measuring stellar reddenings using machine learning predictions of intrinsic colors, analogous to the traditional pair method. The model's predictions are based on a set of 1.5M low-reddening ($\ebv < 0.04$) stars, whose target variable is the observed color dereddened using SFD and the features are DESI DR2 spectroscopic parameters (\teff, \feh, \logg). We use this model to predict intrinsic colors for a set of 3.3M stars observed by DESI, intending to include as many stars as possible while eliminating target classes and objects with potentially unreliable measurements. We then subtract these from the observed colors to produce reddenings in $g-r$, $r-z$, $z-W1$, $W1-W2$, $g-W1$, and $g-W2$ colors. The small-scale residual fluctuations in our model's predictions are 17~mmag in $g-r$ and 77~mmag in $g-W1$.

We use the synthetic MARCS grid spectrum to produce a relation mapping the ratio of reddenings ${E(g-W1)/E(g-r)}$ to $R(V)$. We then apply this model to our measured reddenings of stars in regions where $E(g-r) > 0.1$ and study the variation in $R(V)$ over that footprint. We draw the following conclusions:

\begin{enumerate}
\item We find a bimodality in the distribution of $R(V)$ against $\ebv$, with peaks at values of 2.9 and 3.4, significant deviations from the canonical average value of 3.1 for the diffuse interstellar medium. These variations introduce errors in extinction corrections that bias the selection of spectroscopy targets for surveys like DESI. The peculiar non-Gaussian distribution is plausibly related to the small number of dust clouds seen in the high-latitude DESI footprint, and makes relying on simple Gaussian priors for the value of $R(V)$ fraught.
\item We find $\sigma_{\rm{NMAD}}(R(V))=0.23$ over the map of our analyzed footprint. The overall trends of high and low $R(V)$ values match the results from the work of ZG24, which is based on \gaia spectrophotometry.
\item The variation in $R(V)$, which we compute using the ratio of reddenings in two colors, is strongly correlated to the ratio of reddening in $g-r$ and the reddening predicted from SFD. Thus, the combination of a single optical reddening with the far-infrared brightness of dust can be used to predict variations in the extinction curve.
\item We do not find good correlations between our measured $R(V)$ and the dust temperature, dust spectral emissivity index $\beta$, or $H\alpha$.
\end{enumerate}

The physical cause of the significant variation in $R(V)$ over the diffuse ISM of high Galactic latitudes mapped by DESI warrants further investigation. Expansion of the approach considered here to larger volumes across the Milky Way disk or into other galaxies may enable the physical drivers to be teased out in the future.  Future work could also correlate maps of $R(V)$ with other tracers of the ISM, such as far-infrared polarization fraction, the ratio of the neutral to ionized gas, or the intensity of the ultraviolet radiation field.

Ongoing and upcoming surveys will make important contributions to dust extinction curve mapping.  Surveys like DESI and the SDSS  \citep{Kollmeier:2017} continue to observe millions of stars in the Milky Way.  {\it Roman} launched in August 2026, and soon the Roman Galactic Plane Survey will enable extinction curve mapping deep into the Milky Way disk \citep{RGPS:2025}.  Hubble observations have enabled extinction curve measurements in M31 \citep{Clayton:2025}, and {\it Webb} observations will be able to extend this to more distant galaxies.  Future work will need to employ these rich data sets to understand the physical drivers of observed variations in the extinction curve in the universe.

\section*{Acknowledgments}
This material is based upon work supported by the U.S. Department of Energy (DOE), Office of Science, Office of High-Energy Physics, under Contract No. DE–AC02–05CH11231, and by the National Energy Research Scientific Computing Center, a DOE Office of Science User Facility under the same contract. Additional support for DESI was provided by the U.S. National Science Foundation (NSF), Division of Astronomical Sciences under Contract No. AST-0950945 to the NSF’s National Optical-Infrared Astronomy Research Laboratory; the Science and Technology Facilities Council of the United Kingdom; the Gordon and Betty Moore Foundation; the Heising-Simons Foundation; the French Alternative Energies and Atomic Energy Commission (CEA); the National Council of Humanities, Science and Technology of Mexico (CONAHCYT); the Ministry of Science, Innovation and Universities of Spain (MICIU/AEI/10.13039/501100011033), and by the DESI Member Institutions: \url{https://www.desi.lbl.gov/collaborating-institutions}.

The DESI Legacy Imaging Surveys consist of three individual and complementary projects: the Dark Energy Camera Legacy Survey (DECaLS), the Beijing-Arizona Sky Survey (BASS), and the Mayall z-band Legacy Survey (MzLS). DECaLS, BASS and MzLS together include data obtained, respectively, at the Blanco telescope, Cerro Tololo Inter-American Observatory, NSF’s NOIRLab; the Bok telescope, Steward Observatory, University of Arizona; and the Mayall telescope, Kitt Peak National Observatory, NOIRLab. NOIRLab is operated by the Association of Universities for Research in Astronomy (AURA) under a cooperative agreement with the National Science Foundation. Pipeline processing and analyses of the data were supported by NOIRLab and the Lawrence Berkeley National Laboratory. Legacy Surveys also uses data products from the Near-Earth Object Wide-field Infrared Survey Explorer (NEOWISE), a project of the Jet Propulsion Laboratory/California Institute of Technology, funded by the National Aeronautics and Space Administration. Legacy Surveys was supported by: the Director, Office of Science, Office of High Energy Physics of the U.S. Department of Energy; the National Energy Research Scientific Computing Center, a DOE Office of Science User Facility; the U.S. National Science Foundation, Division of Astronomical Sciences; the National Astronomical Observatories of China, the Chinese Academy of Sciences and the Chinese National Natural Science Foundation. LBNL is managed by the Regents of the University of California under contract to the U.S. Department of Energy. The complete acknowledgments can be found at \url{https://www.legacysurvey.org/}.

Any opinions, findings, and conclusions or recommendations expressed in this material are those of the author(s) and do not necessarily reflect the views of the U. S. National Science Foundation, the U. S. Department of Energy, or any of the listed funding agencies.

The authors are honored to be permitted to conduct scientific research on I'oligam Du'ag (Kitt Peak), a mountain with particular significance to the Tohono O’odham Nation.

Sarv Parteek Singh acknowledges support from Space Telescope Science Institute's Director's Research Fund under project number D0001.82533. Sergey Koposov acknowledges support from Science \& Technology Facilities Council (STFC) (grant ST/Y001001/1).

Generative AI tools were used in the preparation of this work. The models used were Claude Opus and Sonnet (Anthropic), GPT-5 and GPT-6 Astra series (OpenAI), and Gemini 3.5 Thinking (Google), the latter accessed through Perplexity. They were used for refinement of the manuscript text, identifying candidate references, drafting code fragments, and discussion and clarification of concepts. All references identified in this way were independently verified against the original publications. No AI-generated content is presented as original research data. All scientific content, analysis, and conclusions are the authors' own, and the authors take full responsibility for the correctness and originality of the content of this paper.

\facilities{Mayall, Blanco, Bok, WISE}

\vspace{2mm}
{\large\textit{Software:}} numpy \citep{numpy:2020}, scipy \citep{scipy:2020}, matplotlib \citep{matplotlib:2017}, astropy \citep{astropy2013a, astropy:2018, astropy:2022}, healpy \citep{healpy:2019, HEALPIX:2005}, scikit-learn \citep{scikit-learn:2011}, dust\_extinction \citep{Gordon_dust_extinction:2024}

\bibliographystyle{aasjournalv7}
\bibliography{extinction}

\end{document}